\documentclass[12pt]{article} 

\usepackage{amsfonts}
 \usepackage{amssymb}
 \usepackage{amsmath}

\def\pmx{\begin{pmatrix}}
\def\emx{\end{pmatrix}}
\def\bsq{\begin{subequations}}
\def\esq{\end{subequations}}

\newtheorem{thm}{Theorem}

\newtheorem{lemma}[thm]{Lemma}

        \def\pf{\medbreak\noindent{\bf Proof:}\enspace}
     \def\rmk{\medbreak\noindent{\bf Remark:}\enspace}

\newcommand{\mE}{\mathcal E}
\newcommand{\mG}{\mathcal G}
\newcommand{\mC}{\mathcal C}

\newcommand{\mW}{\mathcal W}
\newcommand{\mU}{\mathcal U}

\newcommand{\kt}[1]{ | #1 \ket }

\newcommand{\bfC}{\mathbf{C}}

\newcommand{\bfZ}{\mathbf{Z}}

 \newcommand{\sa}[2]{\scriptsize{ \begin{array}{c} #1 \\ #2 \end{array}}}

\def\be{\begin{eqnarray}}
\def\ee{\end{eqnarray}}
\def\bee{\begin{eqnarray*}}
\def\eee{\end{eqnarray*}}
\def\ds{\displaystyle}
\def\nl{\newline}
\def\ts{\textstyle}

\def\ovb{\overline}

\def\bra{\langle}
\def\ket{\rangle}

\def\iff{\Leftrightarrow}

\def\raw{\rightarrow}

\def\half{{\textstyle \frac{1}{2}}}
\def\1rt2{{\textstyle \frac{1}{\sqrt{2}}}}

\def\ot{\otimes}

\def\wh{\widehat}
\def\wtd{\widetilde}
\def\vep{\varepsilon}

\def\re{\, {\rm Re} \, }
\def\im{ \, {\rm Im} \, }

\def\ds{\displaystyle}
\def\ts{\textstyle}
\def\nl{\newline}

\def\nn{\nonumber}
\def\wt{{\rm wt}}
\def\sp{{\rm s}}

\def\mm{ \! - \!}

\begin{document}

\title{Permutationally
Invariant Codes  \\ for Quantum Error Correction}

\author{Harriet Pollatsek \\ Department of Mathematics and Statistics \\ 
  Mount Holyoke College \\ South Hadley, Massachusetts 01075 
  \\ {\normalsize hpollats@MtHolyoke.edu}
\and Mary Beth Ruskai\thanks{The work of MBR was partially supported  by
 the National Security Agency (NSA) and
 Advanced Research and Development Activity (ARDA) under
Army Research Office (ARO) contract number 
     DAAD19-02-1-0065, and by the National Science
        Foundation under Grant  DMS-0314228.} 
   \\ Department of Mathematics \\
Tufts University\\
  Medford, Massachusetts 02155 \\ {\normalsize marybeth.ruskai@tufts.edu}}


\maketitle

\begin{abstract}
A permutationally invariant n-bit code for quantum error correction can
be realized as a subspace stabilized by the non-Abelian group $S_n$.
The code is  spanned by bases for the trivial representation, and all
other irreducible representations, both those of higher dimension and orthogonal
bases for the trivial representation, are available for error correction.

A number of new (non-additive) binary codes are obtained, including two
new 7-bit codes and a large family of new 9-bit codes.   
It is shown that the degeneracy
arising from permutational symmetry facilitates the correction of
certain types of two-bit errors.   The correction
of two-bit errors of the same type is considered in detail, but 
  is shown not to be compatible with single-bit error
correction using 9-bit codes.  
\end{abstract}

\pagebreak

\tableofcontents

 \pagebreak

\section{Introduction}  \label{sect:intro}  


Quantum error correction is now well-developed in the case of
those stabilizer codes \cite{CRSS1,G1}, which arise as subspaces 
stabilized by
Abelian subgroups of the Pauli group.   These codes,
also known as ``additive codes,'' can be regarded as an extension
of classical binary codes over $Z_2$ to  codes over $GF(4)$ which
satisfy an  additional orthogonality condition.   They 
generalize  the classical notion of distance
and thus seem best suited to situations in 
which all one-bit errors are equally likely and the noise is
uncorrelated. 

There are other approaches to fault tolerant computation which 
use structures which are resistant to decoherence, e.g.,
topological quantum computation and decoherence free (DF)
subspaces or  subsystems.   (See \cite{Ktv} and \cite{LW}
respectively for further discussion and references.)  Some
physical implementations may also be designed to protect
against certain types of errors.   Much of the current 
analysis is based on simple models using independent errors. 
In more realistic models  some types of correlated errors may
be more probable than arbitrary two-bit errors (and possibly 
even than certain one-bit errors).   Hybrid approaches to fault
tolerance which combine resiliency (either through encoding
or hardware design) with error correction may require codes
with properties different from those stabilized by Abelian
subgroups of the Pauli group.

It is now known \cite{RHSS} that other types of quantum codes, often
called ``non-additive,''  exist.    Although some attempts \cite{RV}
have been made to develop classes of non-additive codes, much
of this work, e.g.,  \cite{Knl,KR1},  has been for non-binary codes. 
In this paper we consider a natural generalization of  stabilizer codes
to binary codes associated with the action of non-Abelian groups.  
  We concentrate our attention on the
symmetric group as a case study, and call a code on which the symmetric
group acts trivially {\em permutationally invariant}.
We will be particularly interested in
the use of higher dimensional representations for the correction of
two-bit errors, and the ways in which the degeneracy associated with
permutational invariance of code words allows the correction of more
two-bit errors than would be expected by simple dimensional arguments.

We find a number of  new codes.  In particular,
we give two new 7-bit codes which are impervious 
to exchange and can correct all one-bit errors together
with some rather special two-bit
errors.   We show that the classical 5-bit repetition code can correct more
two-bit quantum errors than those associated with a single type of one-bit
error.   We show that there is a large family of permutationally invariant
9-bit codes in addition to the simple one found in \cite{Paul1}.
Unfortunately, none of these 9-bit codes is as powerful for two-bit error
correction as one might expect.

Although the discovery of new codes is always of interest, we emphasize
that our primary goal is to study permutationally invariant codes as 
{\em   examples of codes obtained from the action of a non-Abelian group}.
These non-Abelian groups will, typically, be more general
than  subgroups of the Pauli group.

It is worth pointing out some significant differences between  
our approach and the ``Clifford codes'' associated with    
``nice error bases'' as proposed by Knill \cite{Knl} and developed
by Klappenecker and R\"otteler \cite{KR1,KR2}.
Their approach, which considers generalizations of the Pauli group
for $d > 2$, yields non-stabilizer codes only for $d \geq 4$;
we obtain new   non-stabilizer codes for $d = 2$.
(Although our approach could, in principle, be applied for any $d$,
 we study only  $d = 2$.)
In the KKR approach, the code is associated with a normal
subgroup $N$ of an error group, but need not come from bases
for the trivial representation of $N$.  
We retain the requirement that a code subspace is spanned
by  bases for the
trivial representation of a group, but the non-Abelian group 
defining our code need not be associated with an error group
in the sense of Knill \cite{Knl}.  From a formal point of view, our
group and error sets reside in an operator algebra associated with
the usual Pauli group, but we do not use this structure.

It was recognized   earlier \cite{BKLW,KBLW}, in
the context of DF subspaces, that
quantum error correcting codes can be obtained as stabilizers for
non-Abelian groups.   However, the use of higher dimensional
irreducible representations for error correction was not explored.
Moreover, the original philosophy underlying the DF approach to fault
tolerant quantum computation, namely, to avoid anything which
might perturb the system out of the stable subspace, is
antithetical to active error correction.    In \cite{BKDLW, KBDW} 
the use of  encoding to facilitate universal computation,
rather than error correction, was introduced.
 Another important development  was the 
generalization of DF subspaces to DF subsystems  \cite{KLV}, in 
which the code itself can transform
as a higher dimensional representation.  
The notion of stabilizer was then modified in \cite{KBLW,Z} to 
encompass DF subsystems  as well.
There is now an extensive literature on  various aspects of 
both DF subspaces and systems, including proposals for
 hybridization of DF methods with active error correction, 
and scenarios in which DF  encoding can replace active error
correction.   We refer the reader to \cite{LW} for references
and further discussion.  

Although motivated by the expected utility of
 codes capable of  correcting specific set of 
correlated errors, we do not present a physical model leading
to such sets.  We deal  only with  construction of codes, leaving
their application within  a full-fledged scheme for fault
tolerance  for further investigation.

The rest of this paper is organized as follows.  In the
next section, we outline the basic set-up and notation we
will use.  We describe different classes of conditions associated
with one-bit errors in Section~3 and analyze them in Section~4.
In Section~5, we consider two-bit error correction.  In
Section~6 we first consider some explicit examples of codes
for $n = 5, 7$ or $9$; we then show that
none of the 9-bit codes can correct all double errors of
one type.   

\section{Preliminaries}

\subsection{Stabilizers and error sets}  \label{sect:err}

In the general situation, we have a set of errors $\mE =
\{  e_1, e_2 \ldots e_M \}$ which we want to correct.  We will also
have a unitary group $\mG$ which acts on the vector space
$\bfC^{2^n}$.  The elements of both $\mE$ and $\mG$ will be linear
operators which act on $\bfC^{2^n}$.  
Typically, these will be non-trivial linear combinations of
elements of the Pauli group, rather than simply tensor products
of Pauli matrices.  In particular, we can consider
$S_n$ as the group generated by the exchange operators $E_{rs}$ which can
be written as
\be   \label{eq:exchg}
  E_{rs} = \frac{1}{2}\left[I \ot I + X_r \ot X_s  + Y_r \ot Y_s  + Z_r \ot
Z_s \right]
\ee
where $X_k, Y_k, Z_k$ denote the action of the $\sigma_x, \sigma_y$ and
$\sigma_z$ operators on bit $k$. 
Note that the set $\{ E_{1s} \, : \, ~ s= 2 \ldots n  \}$ suffices to
generate the group $S_n$.

Since $\bfC^{2^n}$ is invariant under the action of $\mG$, it can be
decomposed into invariant subspaces corresponding to irreducible
representations of $\mG$.  As is well-known \cite{Ham,S}, those subspaces
corresponding to inequivalent representations are orthogonal, and those
for equivalent representations can be chosen orthogonal.   We 
want to exploit the freedom in the latter 
to construct codes with particular properties, and use the additional
orthogonality from inequivalent representations for error correction.

 Now suppose
there is a subspace $T$ which is stabilized by
$\mG$, in the sense $g | w \ket =  | w \ket $ for all $g \in \mG$ and all 
$|w \ket \in T$.
Consider a
subset of errors $\mE'$ which is  invariant under $\mG$ in the
sense $g e_p g^{-1} \in \mE'$ for $e_p \in \mE'$.  Then the space $\mE'(T)$
spanned by
$\{ e_p | w \ket  \, : \, e_p \in \mE', ~ | w \ket \in T \}$
is also invariant under $\mG$ since
\be
   g e_p | w \ket = (g e_p g^{-1}) g | w \ket = e_q | w \ket .
\ee
Hence the space $\mE'(T)$ can be decomposed into an orthogonal sum
corresponding to irreducible representations of $\mG$.
Since the span of $\mE'$ itself is invariant under
 $\mG$, it too can  be
decomposed  into a sum of irreducible subspaces. In fact, one can regard the
two spaces $\mE'(T)$ and span$\{ \mE' \}$ as being decomposed in parallel
into orthogonal sums corresponding to
irreducible representations of $\mG$.

For example, the set of single bit flips
$\mE'_X =\{ X_1, X_2, \ldots X_n  \}$
is invariant under $S_n$.  In fact,
its span is isomorphic to the standard
$n$-dimensional representation of $S_n$, which decomposes into the sum
of the trivial  representation, spanned by $\sum_r X_r$, and the
$(n\!-\!1)$-dimensional irreducible representation,  spanned by  
${X_1 - X_2}, \ldots, X_1 - X_n$.    Similar considerations hold for
the errors $\{ Y_1,  \ldots Y_n  \}$, and $\{ Z_1,  \ldots Z_n  \}$

The resulting linear combinations of errors  $X_p - X_q$
 may not be invertible.  However, this poses no problems for error correction
because we will only need to ``invert"  when a measurement shows
we are in $(X_p - X_q) T$ which is orthogonal to the null space
of $(X_p - X_q)$.

In view of their  role as bases for the trivial
representation, it is useful to define the average errors 
$\ovb{X}, \ovb{Y}, \ovb{Z}$ as
\be
 \ovb{X} = \frac{1}{n} \sum_{k = 1}^n X_k  ~~~~
  \ovb{Y} = \frac{1}{n} \sum_{k = 1}^n Y_k ~~~~
  \ovb{Z} = \frac{1}{n} \sum_{k = 1}^n Z_k
\ee
Note that
  $X_r \mm X_s = (X_1 \mm X_s) - (X_1 \mm X_r)$, 
and recall that a code that can correct  errors in a set ${\cal E}$ can also
correct any complex linear combination  of these errors.  Thus, the
error sets
\be  \nn
{\cal E} & = & \{I, X_k, Y_k, Z_k,  k = 1  \ldots n \}, ~~\hbox{and} \\
{\cal E} & = & \{I, \, \ovb{X}, \, \ovb{Y}, \, \ovb{Z}, \, 
  X_1 \mm X_k, \, Y_1 \mm Y_k, \, Z_1 - Z_k, \, k = 2  \ldots n \}  \label{eq:errset}
\ee 
are equivalent.

\subsection{Notation}

The $2^n$dimensional complex vector space $\bfC^{2^n}$ has an orthonormal
basis $\{ |v \ket = |v_1, v_2 \ldots v_n \ket | \}$ indexed by binary $n$-tuples 
$v = (v_1,v_2 \ldots v_n ) \in (\bfZ_2)^n$.   If an  orthonormal basis
$  |0 \ket, |1\ket  $ for $\bfC^2$ is fixed,  this is simply the basis of tensor
products of the form $|v_1 \ket \ot |v_2 \ket \ot \ldots \ot |v_n \ket$
with each $v_i \in \bfZ_2$. The symmetric group
$S_n$ acts on $\bfC^{2^n}$   via a natural
action on these basis vectors; if ${\cal P}$ takes 
$(1 \ldots n) \mapsto (i_1 \ldots i_n)$, then 
${\cal P} |v_1, v_2 \ldots v_n \ket = |v_{i_1}, v_{i_2} \ldots v_{i_n} \ket$.
Define
\be
  {\cal W}_k = \mbox{span} \{\,| v \ket \, : \, \wt(v)=k \, \}
\ee 
where $\wt(v)$  is the number of $k$ for which $v_k = 1$.  
(This is the classical   Hamming  weight of  $v$.) 
Then $\bfC^{2^n} = \bigoplus_{k = 0}^n \mW_k$ is the orthogonal direct sum of the
$\mW_k$.  Moreover, each $\mW_k$ is  invariant under $S_n$ 
and can  be
further decomposed into an orthogonal sum of spaces affording inequivalent 
irreducible representations of $S_n$.  This yields an orthogonal 
decomposition of $\bfC^{2^n}$ into irreducible subspaces.
However, unlike the regular representation, some irreducible representations
occur more than once in $\bfC^{2^n}$, and others not at all.
Appendix~\ref{sect:irred} describes  the 
decomposition of $\mW_k$  into irreducible subspaces for $n = 5, 7, 9$.

Each $\mW_k$ contains the trivial representation, for
which we introduce the basis vector
\be  \label{eq:Wdef}
W_k = \sum_{\wt(v)=k} |v\ket =  \sum_{\cal P} {\cal P}|\underbrace{ 1  \ldots 1}_\kappa
          \underbrace{ 0 \ldots 0}_{n-\kappa} \, \ket
\ee
where the second sum ranges over those permutations ${\cal P}$ which
yield distinct vectors $|v\ket$.  Thus
$\bra W_k , W_k \ket = \binom{n}{k}$.  Occasionally we will
use the normalized vectors $\wh{W}_k  = \binom{n}{k}^{-1/2} W_k$.
Although normalized vectors are useful for many purpose,  those
denoted $W_k$ are more convenient in combinatoric computations.

Finally, we will make repeated use of the combinatoric identity
\be  \label{eq:comb}
  \binom{N}{K} - \binom{N}{K - J} = \frac{N-2K+J}{N+J} \binom{N+J}{K} 
\ee
which  holds  for  J = 1,2  and is easy to verify.  We will occasionally
use the convention that $\binom{N}{K} = 0$ when $N < K$.

\subsection{Codes}  \label{sect:code}

Given a (possibly non-Abelian) group $\mG$, we define a code 
$\mC$ as a subspace 
 of   $\bfC^{2^n}$ which is stabilized by $\mG$ in the sense
\be
     g | v \ket = |v \ket  \quad \mbox{ for all } g \in \mG  ~\hbox{and all} ~ 
        | v \ket \in \mC .
\ee 
If $\mC$ has dimension $2^m$, then one can effectively encode $m$
logical binary units in $n$ physical qubits.  We will restrict ourselves
here to the simple case of $1$ to $n$ encoding, 
for which $m = 1$ and $\mC$ is two-dimensional.  
A code is often specified by  an orthonormal basis for $\mC$,
in which case each basis vector, or ``code word'' can be regarded as a basis for
 the trivial representation of $\mG$.
In the case of two-dimensional codes, we can interpret these basis vectors 
as a logical $0$ and $1$, and will label them $|c_0 \ket$ and $|c_1 \ket$
accordingly.


We now consider two-dimensional codes for the group
$\mG=S_n$. 
If  $|v \ket = |v_1, v_2 \ldots v_n \ket$ is a basis vector of ${\bf C}^{2^n}$
of weight $k$, (or, equivalently, a
binary n-tuple of weight $k$) then $\{ g |v \ket : g \in S_n \}$  is the set of
all basis vectors of weight $k$.  Therefore, any  vector
satisfying 
$g | v \ket = |v \ket  ~ \mbox{ for all } g \in  S_n$ must have 
the form  $\sum_k  a_k W_k$,
 so that we can write a {\em
permutationally invariant code} as a pair of basis vectors of the form
 \be  \label{eq:code.gen}
|c_0 \ket = \sum_k a_k W_k \quad \mbox{ and } \quad |c_1 \ket = \sum_k b_k
W_k 
\ee
for some complex numbers $a_k$ and $b_k$ with $\sum_k \ovb{a}_k b_k = 0$.
 A vector $\sum_k
d_k W_k$ is called {\em even} (resp. {\em odd}) if $d_k$ is nonzero only 
when $k$ is even (resp. odd).

Note that we have defined a code so that the individual basis vectors
are permutationally invariant.   This is a
stronger requirement than that the subspace defined by the code is
invariant under $S_n$.   However, the distinction is unlikely to 
matter in practice.
In the case of two-dimensional codes, the two types of
invariance are equivalent  whenever $n > 3$.   In general
one can have an invariant subspace of dimension $2^m$ only if it
can be written as a direct sum of irreducible subspaces whose  dimensions
sum to $2^m$;  this will usually consist of  $2^m$ copies
of the trivial representation, in which case the code words are
also invariant.


We will be primarily interested in codes of the form (\ref{eq:code.gen})
which also satisfy the following two conditions (which together imply
that $n$ is odd).
\begin{itemize}

\item[I)] $b_k = a_{n-k}$ or, equivalently, $|c_1 \ket = (\otimes_j X_j)  |c_0 \ket$.

\item[II)]  $c_0$ is even and $c_1$ is odd  or,
equivalently,    $ ( \otimes_j Z_j ) |c_{\ell} \ket = (-1)^{\ell}  |c_{\ell}  \ket$.

\end{itemize}
When (I) and (II) both hold, we can write
\be \label{eq:code.spec}
  |c_0 \ket = \sum_{j = 0}^{(n-1)/2}
a_{2j} W_{2j}, \hskip1cm
     |c_1 \ket = \sum_{j = 0}^{(n-1)/2} a_{n-2j-1} W_{2j+1} .
\ee
In addition to simplifying the analysis and ensuring that certain inner
products are zero, these assumptions serve another purpose.  They
ensure that the logical $X$ and $Z$ operations can be implemented
on the code words by $\otimes_j X_j$ and $\otimes_j Z_j$ 
respectively.   Since the actual use of codes in
fault tolerant computation requires
a mechanism for implementing gates on the code words \cite{G2}, this is an
important consideration.   Moreover, there is little loss of
generality in this assumption.   The operators $\otimes_j X_j$
and $\otimes_j Z_j$  lie in the commutant of $S_n$.  Therefore,
they necessarily map invariant subspaces of $S_n$ to invariant
subspaces of $S_n$.   In the case of the code space, we require
the stronger condition that 
\be \label{eq:logic}
[\otimes_j X_j] \mC = \mC ~~~\hbox{and} ~~~[ \otimes_j Z_j] \mC = \mC.
\ee
When (\ref{eq:logic}) holds, there is no loss of generality in assuming
(I) and (II).  These simply restrict the choice of basis in  way that
is convenient and can always be satisfied.

Our goal is to construct a permutationally invariant 2-dimensional code
that can correct  all single qubit errors, and to examine the types
of two-bit errors that can be corrected.

As noted at the end of Section~\ref{sect:intro}, non-Abelian stabilizers 
were considered previously in the context of DF subspace codes.   
Conversely, one can consider a permutationally
 invariant code as a DF subspace which arises from the
highly idealized  situation in which a quantum computer is completely
insulated from its environment, but the  qubits are the spin components of
identical particles which interact.\footnote{For a complete analogy to DF  
subspaces, only those particles within each logical encoded unit 
would be permitted to interact with each other.   Interactions between
particles in different units, would lead to exchange errors
between units.  However, since these would appear as single
bit errors in each logical unit, they would also be correctable.}  
Then, as discussed in
\cite{Paul1}, even in the absence of spin-spin interactions, the Pauli principle
induces an effective interaction between qubits whose DF subspace group 
is precisely that
generated by exchanges.   In essence, the Pauli principle requires 
correlations between the spatial and spin components so that
spatial interactions (such as the Coulomb interaction)  
affect the spin components.  The result of tracing over the
spatial component yields a completely positive map on the spin
components, as in the standard noise model.  Only  a
fully symmetric spin function allows the full wave function to be a product 
(with an anti-symmetric spatial function) consistent with the Pauli 
principle for fermions.   Thus, a DF subspace is precisely one which transforms
as the trivial (or fully symmetric) representation of $S_n$. 
Although this is not a very realistic DF  scenario, it
is useful to see how codes constructed for different purposes 
can be interpreted within the DF subspace , as well as the stabilizer, 
formalism.


\section{Error correction conditions}  \label{sect:basic}

The now well-known necessary and sufficient condition \cite{BDSW,KL1} 
for the code
${\cal C}$ to correct errors in a set ${\cal E} = \{ e_1 \ldots e_M \}$ 
can be stated as
\be \label{KL}
\bra e_p c_i  ,  e_q c_j \ket = \delta_{ij} d_{pq}  
     \quad \quad \forall ~~ e_p, e_q ~ \in ~ \mE. 
\ee
where the matrix $d_{pq}$ does not depend on $i,j$.
One often chooses codes for which $d_{pq} = \delta_{pq} \mu_p$,
but that is not necessary.   Indeed, the   requirement
\be
d_{pq} \equiv
  \bra e_p c_0  ,  e_q c_0 \ket = \bra e_p c_1  ,  e_q c_1 \ket,
\ee
which is implicit in (\ref{KL}), implies that one can always transform
the error set into a modified one $\widetilde{\mE}$ for which the stronger
condition
$\widetilde{d}_{pq} = \delta_{pq} \mu_p$ holds.

Strictly speaking one can only determine whether or not a particular
set of errors is correctable; not whether a particular error
or type of error is ``correctable''.  However, it is often natural
to look for codes for which the set of correctable errors
includes all errors of a particular
type, e.g., the one-bit errors.  One can then ask what additional
errors could be added to this subset to yield a set $\mE$
satisfying (\ref{KL}).   In our discussion of such
situations, the subset involved may be implied by the context.

We will find it useful to think of (\ref{KL}) as defining a pair of
matrices $D^{ii}$ with elements
  $d_{pq}^i = \bra e_p c_i  ,  e_q c_i \ket$ for $i = 0,1$
and a matrix $B = D^{01}$ with elements
$b_{pq} \equiv d_{pq}^{01} = \bra e_p c_0  ,  e_q c_1 \ket$.
Then  (\ref{KL}) is equivalent to the requirements
 $B = 0$  and $D^{00} = D^{11}$. (Because 
  $d_{pq}^{10}$ and $d_{qp}^{01}$ are complex conjugates,
$D^{01} = 0  \iff D^{10} = 0$.  Hence we need not
consider $D^{10}$ explicitly and
will use only $B \equiv D^{01}$.)  For simplicity,
we omit the superscript in $d_{pq}^i$.


For example, the 9-bit permutationally invariant code in \cite{Paul1}  corrects
single qubit errors as well as the  Pauli exchange errors (transpositions)
$E_{rs}$,  for the 36 unordered pairs $r,s$.  We can consider the above
matrices with respect to the errors
$\mE = \{I, E_{rs} , X_1, \ldots, X_9, Y_1, \ldots, Y_9, Z_1, \ldots, Z_9\}$.
It was shown that
$D^{00} = D^{11}$  and has the block diagonal form
\be  \label{eq:D1blk}
\pmx D_0 & 0 & 0  & 0 \\  0 & D_{XX} & 0  & 0 \\
   0 & 0 & D_{YY}  & 0 \\   0 & 0 & 0  & D_{ZZ} \emx
\ee
where $D_0$ is a  $37 \times 37$ rank one matrix and the $9 \times 9$
matrices
$D_{XX}, D_{YY}, D_{ZZ}$  correspond to the one-bit errors indicated
by the subscripts.  These all have the cyclic form
\be  \label{eq:cycblk}
   \pmx a & b &   \ldots & b \\  b & a &   \ldots & b \\
    \vdots &  ~ & ~ & \vdots \\ b & b &   \ldots & a \emx .
\ee

For {\em any} permutationally invariant code the blocks $D_{XX}, D_{YY},
D_{ZZ}$
necessarily have the form (\ref{eq:cycblk}).  Such matrices can always be
diagonalized by a change of basis to $(1, 1 \ldots 1)$ and its orthogonal
complement.  This corresponds to replacing the errors
$\{ f_1, f_2 \ldots  f_n \}$ by the corresponding average
 $\{ \ovb{f} \}$ and a suitable orthogonalization of
$\{ f_1 - f_k , k = 2 \ldots n \}$ where $f$ denotes any of $X, Y, Z$.

Now the orthogonality of subspaces associated with different
irreducible representations ensures that
\be  \label{eq:errortho}
  \bra \ovb{f} W_j  ,  (g_r - g_s) W_k \ket = 0
\ee
for all $j,k$ and any choice of $\ovb{f}= I, \ovb{X} , \ovb{Y}, \ovb{Z} $ and 
$g = X, Y, Z$. Alternatively, we
 can show this directly by observing that the exchange operator
$E_{rs}$ is unitary so that
\bee
  \bra \ovb{f} W_j  ,  (g_r - g_s) W_k \ket & = &
    \bra E_{rs} \ovb{f} E_{rs}(E_{rs}W_j) ,  E_{rs} (g_r - g_s)E_{rs}
(E_{rs} W_k) \ket \\
  & = & \bra \ovb{f} W_j  ,  (g_s - g_r) W_k \ket \\
   & = & - \bra \ovb{f} W_j  ,  (g_r - g_s) W_k \ket
\eee
which implies (\ref{eq:errortho}).  For such codes,
each of the matrices $D^{ii},  (i = 0,1)$ and $B$ have the form below
(which we write only for $D$) with respect to the order in (\ref{eq:errset}).
\be  \label{eq:Dform}
   \pmx \begin{array}{cccc}
   d_{II} & d_{IX} & d_{IY} & d_{IZ} \\
    d_{XI} & d_{XX} & d_{XY} & d_{XZ}  \\
  d_{YI} & d_{YX} & d_{YY} & d_{YZ} \\
  d_{ZI} & d_{ZX} & d_{ZY} & d_{ZZ}   \end{array}  & {\LARGE {\mathbf 0}}  \\ {\LARGE
  \mathbf 0}  &
    \begin{array}{ccc}
     D_{XX} & D_{XY} & D_{XZ}  \\
  D_{YX} & D_{YY} & D_{YZ} \\
 D_{ZX} & D_{ZY} & D_{ZZ}  \end{array} \emx
\ee

Conditions (I) and (II) immediately give many additional zero entries.
One nice way to see which entries are zero is to
observe that  $\ot_k Z_k$ commutes with $Z_r$ and
anti-commutes with
$X_r$ and $Y_r$ for all $r$.   Thus, for every one-bit error $e_p$, 
 $e_p (\ot_k Z_k) = \epsilon_p^Z (\ot_k Z_k) e_p$, where
\be
\epsilon_p^Z = \left\{ \begin{array}{lll}
  +1 &   \mbox{  for  } & e_p \in \{I, \ovb{Z}, (Z_r - Z_s) \}    \\
 -1 & \mbox{  for  }  & e_p \in \{ \ovb{X}, \ovb{Y}, (X_r - X_s), (Y_r -
Y_s) \}
  \end{array} \right.
\ee
Also, $\ot_k Z_k$  is unitary so that
\be
\bra e_p c_i  ,  e_q c_j \ket &=& \bra   (\ot_k Z_k) \, e_p c_i  , 
     (\ot_k Z_k) \, e_q c_j \ket \nn \\
&=& \epsilon^Z_p \epsilon^Z_q
   \bra e_p \, (\ot_k Z_k) c_i  ,  e_q \, (\ot_k Z_k) c_j \ket \nn \\
&=& \epsilon^Z_p \epsilon^Z_q  (-1)^{i+j}
   \bra e_p c_{i }  ,  e_q c_{j } \ket. \nn
\ee
From this we can conclude the following.  
\begin{itemize}

\item[A)]  When $i = j$,
$\bra e_p c_i  ,  e_q c_i \ket = 0$ whenever $\epsilon^Z_p \neq \epsilon^Z_q$
Thus, $d_{IX} = d_{IY} = d_{XZ} = d_{YZ} = 0$ and $D_{XZ} = D_{YZ} = 0$.

\item[B)]  When $i \neq j$,
  $\bra e_p c_i  ,  e_p c_j \ket = 0 $  whenever $\epsilon^Z_p = \epsilon^Z_q$,
from which we can conclude $b_{IZ}=b_{ZI}=b_{XY}=b_{YX}=0$ and
$B_{XX}=B_{YY}=B_{ZZ}=B_{XY}=B_{YX}=0$.

\end{itemize}

Combining this with $d_{fg}=0 \Leftrightarrow d_{gf}=0$ and 
$D_{fg}=0 \Leftrightarrow D_{gf}=0$, we find that the $D^{ii}$ have the form
\be  \label{eq:blkii}
    \pmx \begin{array}{cccc}
   d_{II} & 0 & 0 & d_{IZ} \\
   0 & d_{XX} & d_{XY} & 0 \\
  0 & d_{YX} & d_{YY} &0 \\
  d_{ZI} & 0 & 0 & d_{ZZ}   \end{array}  & {\LARGE   0}  \\
  {\LARGE   0}  &
    \begin{array}{ccc}
     D_{XX} & D_{XY} & 0 \\
  D_{YX} & D_{YY} & 0 \\
 0 & 0 & D_{ZZ}  \end{array} \emx
\ee
and $B$ has the form
\be \label{eq:blk01}
   \pmx \begin{array}{cccc}
   0 & b_{IX} & b_{IY} & 0 \\
    b_{XI} & 0 & 0 & b_{XZ}  \\
  b_{YI} & 0 & 0 & b_{YZ} \\
  0 & b_{ZX} & b_{ZY} & 0  \end{array}  & {\LARGE \mathbf 0}  \\
  {\LARGE \mathbf 0}  &
   \begin{array}{ccc}
      0 & 0 & B_{XZ}  \\
   0 & 0 & B_{YZ} \\
 B_{ZX} & B_{ZY} & 0  \end{array} \emx
\ee

Now, observe that  $\ot_k X_k$ commutes with $X_r$ and
anti-commutes with  $Y_r$ and $Z_r$.  Proceeding as above,
we find
\be   \label{eq:errXgp}
\epsilon_p^X = \left\{ \begin{array}{lll}
  +1 &   \mbox{  for  } & e_p \in \{I, \ovb{X}, (X_r - X_s) \}    \\
 -1 & \mbox{  for  }  & e_p \in \{ \ovb{Y}, \ovb{Z}, (Y_r - Y_s), (Z_r -
Z_s) \}
  \end{array} \right.
\ee
and\be
\bra e_p c_i  ,  e_q c_j \ket &=& \bra (\ot_k X_k) \, e_p c_i  ,  
   (\ot_k X_k) \, e_q c_j \ket \nn \\
&=& \epsilon^X_p \epsilon^X_q \bra e_p \, (\ot_k X_k) \, c_i  ,  e_q \, (\ot_k X_k) \, c_j
\ket \nn \\
&=& \epsilon^X_p \epsilon^X_q \bra e_p c_{i+1}  ,  e_q c_{j+1} \ket, \nn
\ee
where we interpret $i+1$ and $j+1$ mod 2.
Thus we can conclude
\begin{itemize}
\item[C)] When $i=j$,  condition (\ref{KL}) holds whenever $\epsilon_p^X =
\epsilon_q^X$.
(This means, in particular, that the diagonal entries and blocks of $D^{00}$
and $D^{11}$ agree.)

\item[D)]  When $i=j$  and $\epsilon_p^X \neq \epsilon_q^X$, condition (\ref{KL})
can only be satisfied if
$\bra e_p c_i  ,  e_q c_i \ket = 0$ for $i=0,1$.  Thus we
must have $d_{IZ} = d_{XY} =  0$ and
$D_{XY} =  0$.

\item[E)]  When $i \neq j$,
$\bra e_p c_0 , e_q c_1 \ket = \pm \bra e_p c_1  ,  e_q c_0 \ket
 = \pm \overline{\bra e_q c_0 , e_p c_1 \ket}$.
Thus, we can conclude, e.g., that matrix entries $b_{XZ}=0 \iff b_{ZX}=0$ and
blocks $B_{XZ}=0 \iff B_{ZX}=0$, so it
suffices to check entries of $B$ above the main diagonal.

\end{itemize}

Thus, when conditions (I) and (II) are satisfied, we find that sufficient
(and necessary)
conditions for (\ref{KL}) to hold are that

\begin{itemize}

\item All off-diagonal entries and blocks in  (\ref{eq:blkii})
are zero,

\item All remaining entries in (\ref{eq:blk01}) are zero.

\end{itemize}
Moreover,  it suffices to check
 matrix elements above the main diagonal in (\ref{eq:blkii}) and
(\ref{eq:blk01}).

We can break these conditions into several groups, which will turn
out to be related or equivalent.
\begin{itemize}

\item[a)] Conditions $b_{IX} = b_{IY} = b_{ZX} = b_{ZY} = 0$, which
are equivalent to \linebreak $\bra c_0  ,  (\ovb{X}\pm i \ovb{Y}) c_1 \ket
  = \bra \ovb{Z}c_0  ,  (\ovb{X} \pm \ovb Y) c_1 \ket = 0$.
These will yield just two conditions when the $a_k$ are real.

\item[b)] Conditions
$d_{IZ} = d_{XY}  = 0$, i.e.,
$\bra c_0  ,  \ovb{Z}c_0 \ket = \bra \ovb{X}c_0  ,  \ovb{Y} c_0 \ket = 0$.
These will reduce to one condition when  the $a_k$ are real.

\item[c)] The block conditions $B_{XZ} = B_{YZ} = 0$, which are equivalent to
\newline
$\bra [(X_1-X_r) \pm i (Y_1 - Y_r)]c_0  ,  (Z_1 - Z_s)c_1 \ket = 0$ for $2
\leq r,s \leq n$.

\item[d)] The block condition $D_{XY}  = 0$, which is equivalent to \nl
$\bra (X_1 - X_r)c_0  ,  (Y_1-Y_s)c_0 \ket = 0$ for $2 \leq r,s \leq n$.

\end{itemize}
We will see that for codes which satisfy conditions (I) and (II) and have
all coefficients real, conditions (c) on blocks will be satisfied
whenever (a) holds; and conditions (d) on blocks will be satisfied
whenever (b) holds.  Thus, we will only need to satisfy three non-linear
equations for such codes.  We can summarize this as follows.


\begin{thm} \label{thm:main}
Assume that the coefficients $a_k$ associated with a 
permutationally invariant code  which has the form
(\ref{eq:code.spec}) and length $n$ are all real.  
Such a code can correct
all one-bit errors if and only if the following equations hold.
\be   \label{eq:crs01.red1}
0 & = & {\scriptsize \frac{n+1}{2}}   \binom{n}{\frac{n+1}{2}}
 a_{\frac{n+1}{2}}^2
 ~ +  2\sum_{m = 1}^{\lfloor(n-1)/4\rfloor} a_{2m} a_{n-2m+1} \,
      2m \binom{n}{2m} \\
 \label{eq:crs01.red2}
0 & = & {\scriptsize \frac{n+1}{2}}   \binom{n}{\frac{n-1}{2}}
a_{\frac{n-1}{2}}^2
 ~ + 2\sum_{m = 0}^{{\lfloor(n-3)/4\rfloor}} a_{2m} a_{n-2m-1}  \, (n-2m)
   \binom{n}{2m}   \\
 0 & = & \sum_{m = 0}^{(n-1)/2} a_{2m}^2 (n-4m) \binom{n}{2m}
    \label {eq:skew2}
\ee
\end{thm}
The theorem will follow from the analysis in the next section.   The
result can be extended to complex $a_k$ as
discussed in Appendix~\ref{app:compx}.   As noted before, it is
implicit in (\ref{eq:code.spec}) that $n$ is odd.


\section{Error condition analysis} \label{sect:anal}

\subsection{Conditions of type (a) --- $c_0, c_1$ orthogonality}
\label{sect:typea}

First we give expressions for the action of the average errors
$\ovb{X}$, $\ovb{Y}$ and $\ovb{Z}$.  It is easy to
check that for $0 \leq k \leq n$
\be
n \ovb{Z} \, W_k &=& (n-2k) W_k  \label{Z1} \\
n \ovb{X} \, W_k &=& (k+1)W_{k+1}+(n-k+1) W_{k-1}
\label{X1} \\
i \, n  \ovb{Y} \, W_k &=& (k+1)W_{k+1} -(n-k+1) W_{k-1},
\label{Y1}
\ee
where it is understood that if $m<0$ or $m>n$, $W_m$ should be
replaced by zero; for example, $n \ovb{X} \,W_0=W_1$.

To analyze the requirement $b_{IX} = b_{IY} = b_{ZX} = b_{ZY} = 0$, it is
equivalent (and somewhat easier) to use the conditions
\be   \label{eq:XY-I} 
  0 & = & \bra c_0 , ( \ovb{X} \pm i \ovb{Y}) \, c_1 \ket, ~~~\hbox{and} ~~~ \\
 0  & = &   \bra \ovb{Z} c_0 , ( \ovb{X} \pm i \ovb{Y}) \, c_1 \ket  \label{eq:XY-Z} 
\ee
When condition (I) holds, these conditions become
\be
n \bra c_0, \, (\ovb{X} + i \ovb{Y})c_1 \ket &=&
2 \sum_{k=1}^{n} \, k \, \binom{n}{k} \, \ovb{a}_k a_{n-k+1}\,\label{sum1a}=0\\
n^2 \bra \ovb{Z}c_0, \, (\ovb{X} + i \ovb{Y})c_1 \ket &=&
2 \sum_{k=1}^n \, (n-2k)k \, \binom{n}{k} \, \ovb{a}_k a_{n-k+1} \, \label{sum2a}=0\\
n \bra c_0, \, (\ovb{X} - i\ovb{Y})c_1 \ket &=&
2 \sum_{k=0}^{n-1} \, (n-k)\, \binom{n}{k} \, \ovb{a}_k a_{n-k-1} \, \label{diff1a} =0\\
n^2 \bra \ovb{Z}c_0 , \, (\ovb{X} - i \ovb{Y})c_1 \ket &=&
2 \sum_{k=0}^{n-1} \, (n-2k)(n-k) \, \binom{n}{k} \, \ovb{a}_k a_{n-k-1}\, \label{diff2a} =0
\ee
Thus far, condition (I) has played a minor role and one can easily
obtain more general conditions by replacing $a_{n-k}$ by $b_k$ above.
Now, however, we make explicit use of the fact that all products
have the form $\ovb{a}_k a_{n-k+1}$ to conclude that the real parts of the 
expressions in (\ref{sum1a}) and (\ref{sum2a}) and in (\ref{diff1a}) and 
(\ref{diff2a}) agree up to sign, which leads to the following lemma. 


\begin{lemma}
When the coefficients $a_k$ are all real,
the equations (\ref{sum1a}) to (\ref{diff2a})
are equivalent in pairs, (\ref{sum1a} )
$\leftrightarrow$ (\ref{sum2a} ) and (\ref{diff1a}) $\leftrightarrow$
(\ref{diff2a} ). When $n$ is odd these reduce to
\be  \label{eq:crs01}
  0 & = &  \sum_{k = 1}^{(n-1)/2} a_k a_{n-k+1}  2k \binom{n}{k} + 
       a_{\frac{n+1}{2}}^2 {\ts \frac{n+1}{2}}   \binom{n}{\frac{n+1}{2}} \\ 
 0 & = &  \sum_{k = 0}^{(n-3)/2} a_k a_{n-k-1} 2(n-k) \binom{n}{k} + 
       a_{\frac{n-1}{2}}^2 {\ts \frac{n+1}{2}}   \binom{n}{\frac{n-1}{2}} . 
\label{eq:crs02}
\ee  
When $n$ is even, similar expressions hold with upper limits of
$n/2$ and $n/2 -1$ respectively but without any extra square terms
analogous to $a_{\frac{n \pm 1}{2}}^2$.
\end{lemma}

When condition (II) holds, equations (\ref{eq:crs01}) and (\ref{eq:crs02})
reduce to (\ref{eq:crs01.red1}) and (\ref{eq:crs01.red2}), which proves the first part of
Theorem~\ref{thm:main}.  

\pf First, 
observe that  for $m \neq \frac{n+1}{2}$, the term $a_ma_{n-m+1}$
occurs twice in (\ref{sum1a}), once for
$m=k$ and once for $m=n-k+1$.  Thus, the
coefficient of $a_ma_{n-m+1}$ is
\bee
m \binom{n}{m} + (n \! - \! m\! +\! 1) \binom{n}{n \! - \! m\! +\! 1}
  = 2m \binom{n}{m}
\eee
The coefficient of the same term in (\ref{sum2a}) is
$$
(n \mm 2m) \, m \, \binom{n}{m} + [n \mm 2(n \mm  m\! +\! 1)]
  \, (n \mm  m\! +\! 1) \,
  \binom{n}{n \mm m\! +\! 1} = -2m \, \binom{n}{m}.
$$
To obtain a general proof and reduction to (\ref{eq:crs01.red1}),  it
 suffices to make the change of variable $k \raw n-k + 1$ when
$k \geq \frac{n+3}{2}$ in (\ref{sum1a} )  and (\ref{sum2a}) and,
as above,
use the elementary identity
\be
m \binom{n}{m} = (n \! - \! m\! +\! 1) \binom{n}{n \! - \! m\! +\! 1} .
  \ee
Similarly,  the change of variable $k \raw n-k - 1$
for $k \geq \frac{n+1}{2}$ in  (\ref{diff1a} ) and
(\ref{diff2a}) yields (\ref{eq:crs01.red2}). ~~ QED


\subsection{Conditions of type (b)--- off-diagonal conditions}

Using  (\ref{Z1}) to (\ref{Y1}) one finds that  
\bee
 \bra n \ovb{X}W_k, \,   i  \, n \ovb{Y} W_k \ket &=& \bra W_k, \, n \ovb{Z} W_k \ket 
  = (n-2k) \binom{n}{k} \\
  \bra n \ovb{X} W_k, \,  in \ovb{Y} W_{k+2} \ket &=& - \bra n \ovb{X} W_{k+2}, \,
     i  \, n\ovb{Y}
W_k \ket  \\
\bra n \ovb{X} W_j, \,  i  \, n \ovb{Y} W_{\ell}\ket &=& 0 \quad \quad \mbox{ if } j-\ell \neq \pm 2, 0.
\eee
From these relations, it follows that 
when all $a_k$ and $b_k$ are real
\be   \label{eq:missed}
n^2  \bra \ovb{X}c_j , i \, \ovb{Y} c_j \ket =
  n  \bra c_j ,  \ovb{Z}c_j \ket =
   \sum_k \alpha_k^2 (n-2k) \binom{n}{k}
\ee
where $\alpha_k$ equals $a_k$ or $b_k$ according as $j$ equals 0 or 1.
Thus, we can conclude that when $a_k, b_k$ are real, the off-diagonal
conditions
$d_{IZ} = d_{XY} = 0$ hold if and only if
\be  \label{eq:skew}
  \sum_k a_k^2 (n-2k) \binom{n}{k} = \sum_k b_k^2 (n-2k) \binom{n}{k} = 0.
\ee
This reduces to to  (\ref{eq:skew2}) when conditions I and (II) are satisfied.  
If some $a_k, b_k$ are not real, then (\ref{eq:missed})
holds with $\alpha_k^2$ replaced by $|a_k|^2$ or $|b_k|^2$, but the
additional condition (\ref{eq:XYcomp}) is needed to ensure that the
imaginary part of
$\bra \ovb{X}c_i ,  i \, \ovb{Y} c_i \ket$ is zero, as  discussed
in Appendix~\ref{app:compx}.


\subsection{Conditions of type (c) --- block 
$c_0, c_1$ orthogonality.}

It will again be useful to replace the separate $X, Y$ equations
by their sums and differences.  The requirement that the blocks
$B_{XZ} = B_{YZ} =  0$    is equivalent to
\be  \label{eq:blk.XY}
    \bra \big[ (X_1 - X_r) \pm i (Y_1 - Y_r) \big] \, c_0 ,  (Z_1 - Z_s)\,
c_1 \ket = 0 .
\ee
for $2 \leq r,s \leq n$.
We now need results from Appendix A.  When conditions (I) and (II) hold, 
equations (\ref{eq:dfZm}), (\ref{eq:dfXYp}) and (\ref{eq:dfXYm}) 
imply that (\ref{eq:blk.XY})
is equivalent to the following pair of equations
\be   \label{eq:XYblk1}
\sum_{m = 0}^{(n-3)/2} \ovb{a}_{2m}  a_{n-2m-1} \binom{n-2}{2m} & = &
    0, ~\hbox{and} \\  \label{eq:XYblk2}
\sum_{m = 1}^{(n-1)/2} \ovb{a}_{2m} a_{n-2m+1} \binom{n-2}{2m-2} & = & 0.
\ee
To see that these are equivalent to
(\ref{eq:crs01.red1}) and (\ref{eq:crs01.red2}),
 again make a change of variable of the form $k \raw n- k\mp 1$ in the second
half of each sum and use the identities
\bee
  \binom{n-2}{k} + (1 \mm \delta_{k0}) \binom{n-2}{n-k-1}  & = &
\binom{n-1}{k} ~ =~
   \frac{ n-k}{n} \,  \binom{n}{k} \\
(1 \mm \delta_{k1}) \binom{n-2}{k-2} +  \binom{n-2}{n-k-1}  & = &
\binom{n-1}{k-1} ~ =~
   \frac{ k}{n} \,  \binom{n}{k} .
\eee

\subsection{Conditions of type (d)--- block off-diagonal conditions}
\label{sect:skew.blk}

We now consider the condition $d_{XY} =   0$ which means that
\be \label{XY9}
   \bra (X_1 - X_r)c_0 ,  i(Y_1 - Y_s) c_0 \ket = 0 
\ee
for all choices of $2\leq r,s \leq n$. The crucial fact is that the 
inner products of this
type with $r = s$ and $r \neq s$ differ only by a factor of $2$ as shown
by (\ref{eq:fact2}) in  Appendix~A.
\begin{thm}  When conditions (I) and (II) are satisfied and the $a_k$ are
real,
(\ref{XY9}) holds if and only if (\ref{eq:skew2}) does.
\end{thm}
\pf 
It follows from (\ref{eq:XVk}) and (\ref{eq:YVk})  that when
    $1,r,s$ are distinct
\be  \label{eq:diagss}
  \bra (X_1 - X_r)c_0 ,  i(Y_1 - Y_r)c_0 \ket
  & = & 2 \bra (X_1 - X_r)c_0 ,  i(Y_1 - Y_s)c_0 \ket
 \nn \\ & ~ & \nn \\
& = &
   2 A_0 + 2 \ds{\sum_{m=1}^{(n-3)/2} \binom{n-2}{2m} A_m}
\ee
where
$A_m= |a_{2m}|^2 - |a_{2m+2}|^2+ a_{2m} \ovb{a}_{2m+2}-\ovb{a}_{2m}a_{2m+2}$.
Thus when the $a_k$ are real,  (\ref{XY9})
will be satisfied for all choices of $r,s$ if
\be  \label{eq:diag}
a_0^2 +   \sum_{m=1}^{(n-3)/2} a_{2m}^2  \left[ \binom{n-2}{2m} -
\binom{n-2}{2(m-1)} 
\right] - (n-2) a_{n-1}^2 = 0.
\ee
One can then conclude that (\ref{eq:diag}) is equivalent to
(\ref{eq:skew2})
if $n \left[ \binom{n-2}{2m} - \binom{n-2}{2(m-1)}  \right]  =
    (n-4m)  \binom{n}{2m}$
which follows from (\ref{eq:comb}) with $N = n-2, K = 2m, J = 2$.

\section{Two-bit errors}

\subsection{Some special types of two-bit errors} \label{sect:spec2}

The standard 7-bit CSS code \cite{NC} can correct two-bit errors of the
form $X_r Z_s$ but not those of the form $X_r X_s$ or $Z_r Z_s$.
The last two are far more likely to occur, especially for nearest
neighbors.   We now consider the effect of two-bit errors of 
the same type, which we call ``double'' errors, on permutationally
invariant codes.

Recall that exchange errors have the form (\ref{eq:exchg}).
Permutationally invariant codes are designed so that exchange errors are
degenerate with the identity, i.e., $E_{rs} |c_j \ket = |c_j \ket$ for $j = 0,1$.
Now consider the following three errors
\bsq \be
  F_{rs} &  = & \half \Big[ I \ot I - X_r \ot X_s  -
       Y_r \ot Y_s  + Z_r \ot Z_s  \Big] \\
  G_{rs} &  = & \half \Big[ I \ot I + X_r \ot X_s  -
       Y_r \ot Y_s  - Z_r \ot Z_s \Big] \\
  H_{rs} &  = & \half \Big[  I \ot I  - X_r \ot X_s
      + Y_r \ot Y_s  - Z_r \ot Z_s \Big]
\ee  \esq
and observe that
\begin{itemize}
 \item  $F_{rs}$ exchanges two bits and multiplies by $-1$ if and only
if the values of the bits are different.
 \item  $G_{rs}$ flips the two bits $r$ and $s$ if and only if they
are the same.
\item  $H_{rs}$ flips the two bits $r$ and $s$ and then multiplies by
$-1$ if and only if they are the same.
\end{itemize}
In a product basis of the form $| 00 \ket, | 01 \ket, | 10 \ket, | 11 \ket$
these operators  are represented by the matrices
\bee
 E_{rs} = \pmx 1 & 0 & 0 & 0 \\ 0 & 0 & 1 & 0 \\ 0 & 1 & 0 & 0
     \\ 0 & 0 & 0 & 1 \emx  & ~ &
  F_{rs} = \pmx 1 & 0 & 0 & 0 \\ 0 & 0 & -1 & 0 \\ 0 & -1 & 0 & 0
     \\ 0 & 0 & 0 & 1 \emx    \\
G_{rs} = \pmx 0 & 0 & 0 & 1 \\ 0 & 1 & 0 & 0 \\ 0 & 0 & 1 & 0
     \\ 1 & 0 & 0 & 0 \emx   & ~ &
  H_{rs} = \pmx 0 & 0 & 0 & -1 \\ 0 & 1 & 0 & 0 \\ 0 & 0 & 1 & 0
     \\ -1 & 0 & 0 & 0 \emx  .
\eee

Any code which can correct all  errors of the type
 $E_{rs}, F_{rs}, G_{rs}, H_{rs}$ can also correct
any error of the form $Z_r Z_s, X_r X_s, Y_r Y_s $,
since an error of one type can be written as a linear
combination of those of the other.
For permutationally invariant codes, these two types of
errors are actually equivalent.
\begin{thm}
If $|\psi \ket$ is permutationally invariant (i.e.,
$E_{rs} |\psi \ket = |\psi \ket $ for all $r,s$), then
the operators $F_{rs}, G_{rs}$ and $ H_{rs}$ have the same effect
 on $|\psi \ket$ as $Z_r Z_s, X_r X_s$, and $Y_r Y_s$
respectively, i.e.,
$F_{rs} |\psi \ket = Z_r Z_s|\psi \ket $,
$G_{rs} |\psi \ket = X_r X_s|\psi \ket $ and
$H_{rs} |\psi \ket =  Y_r Y_s|\psi \ket $.
\end{thm}
\pf First note that $E_{rs} + F_{rs} = I + Z_r Z_s$.
Then
\bee
F_{rs} |\psi \ket = \big( I + Z_r Z_s - E_{rs} \big)\, |\psi \ket
   =  |\psi \ket + Z_r Z_s|\psi \ket - |\psi \ket = Z_r Z_s|\psi \ket
\eee
The other two cases are done similarly using $E_{rs} + G_{rs} $
and $E_{rs} + H_{rs} $ respectively.

\subsection{Two-bit error correction conditions}  \label{sect:twobt.cond}

We begin with some simple, but fundamental, results.  
The first follows from the fact that all double errors preserve parity.
\begin{thm} \label{thm:doub.par}
Whenever condition (II) is satisfied,
\be
   \bra e_p c_0 , e_q c_1 \ket = 0
\ee
for any pair of errors in the set $\{ I, Z_r Z_s, X_r X_s, Y_r Y_s\}$
or, equivalently, in the set \linebreak
$\{ I, E_{rs}, F_{rs}, G_{rs}, H_{rs} \}$.
\end{thm}
The next theorem says that all inner products of the form
$\bra Z_rZ_s c_j , Z_q Z_t \, c_j \ket$, $ \bra Z_rZ_s c_j , X_q X_t \, c_j \ket$
etc. are independent of $j = 0,1$.    It follows easily from
the equivalence of condition (I) to
$(\ot_k X_k) \, |c_0 \ket = |c_1 \ket$, and  the fact that $\ot_k X_k$
is a unitary operator which commutes  with any error of the form
$X_r X_s$, $Y_r Y_s$ or $Z_r Z_s$. 
\begin{thm}  \label{thm:doub}
Whenever condition (I) is satisfied,
\be
  \bra f_rf_s c_0 , g_q g_t \, c_0 \ket = \bra f_rf_s c_1 , g_q g_t , c_1 \ket
\ee
where $f, g$ denote any of $\{ X, Y, Z\}$ (the same as well as different)
and $r,s,q,t$ are arbitrary.
\end{thm}

One is  often interested in knowing which two-bit errors can
be  corrected in addition to one-bit errors.  
Conditions involving the average error $\ovb{ZZ}$ can be
readily calculated  by noting that
\be
 \sum_{r \neq s} Z_r Z_s = \sum_{r, s} Z_r Z_s - \sum_r Z_r^2
    = \Big( \sum_r Z_r \Big)^2 - nI = (n \ovb{Z})^2 - nI .
\ee
Combining this with (\ref{Z1}), one finds
\be  \label{eq:doubZ}
 n(n-1) \ovb{ZZ} W_k  =
  \big( \sum_{r \neq s} Z_r Z_s \big) \, W_k  =
     [(n-2k)^2 - n] \,  W_k   .
\ee
 The additional conditions
needed to correct all errors of the form $Z_r Z_s$ include 
$\bra \ovb{Z}c_0, \ovb{ZZ}c_0 \ket = 0 $ 
and $\bra \ovb{ZZ} c_0 , ( \ovb{X} \pm i \ovb{Y}) \, c_1 \ket  = 0$.
The latter gives the following pair of conditions
\be
0 & = & 2 \sum_{k=1}^n \, [(n-2k)^2 - n]k \, \binom{n}{k} \, \ovb{a}_k a_{n-k+1} \,
\label{XYp.ZZ} \\  0 & = &
2  \sum_{k=0}^{n-1} \, [(n-2k)^2 - n](n-k) \, \binom{n}{k} \, \ovb{a}_k a_{n-k-1}\,
\label{XYm.ZZ}  
\ee
Using (\ref{Z1}) and  (\ref{eq:doubZ}), one finds that
$\bra \ovb{Z}c_0, \ovb{ZZ}c_0 \ket = d_{Z,ZZ} = 0 $ is equivalent to
\be
 0 &= & 
\sum_k |a_k|^2 (n-2k) [(n-2k)^2 - n] \binom{n}{k} \label{eq: Z,ZZ}.
\ee

Although one can write down a similar set of conditions for the
correction of errors of the form $X_rX_s$, it is probably easier to use
the following observation.
\begin{thm}  \label{thm:XZequiv}
A permutationally invariant code $|c_0 \ket, |c_1 \ket$ which satisfies
conditions  (I) and {\em(II)  [and corrects a specified set  of one-bit errors] }
can correct all errors of the form $Z_rZ_s$ if and only if the code
\be  \label{eq:Had}
   |C_j \ket = H^{\ot n}|c_0 \ket + (-1)^j H^{\ot n}|c_1 \ket
\ee
can correct all errors of the form $X_rX_s$.
\end{thm}
The map  $|c_j \ket \mapsto |C_j \ket$ in (\ref{eq:Had}) consists
of a Hadamard gate acting on all qubits, followed by an
effective  Hadamard operation
on the resulting code words themselves.  This is extremely useful and
is its own inverse.  We will refer to it as the ``Hadamard code map''.

\subsection{Degeneracy enhancement of classical codes}  \label{sect:deg}

It follows from Theorems \ref{thm:doub.par} and  \ref{thm:doub} that {\em every}
permutationally invariant code for which both conditions (I) and (II) are 
satisfied can correct {\em all} double errors of the form 
$ Z_r Z_s, X_r X_s, Y_r Y_s$ provided that we do not also require 
 single bit errors to be correctable.   
For example, the 3-bit repetition code $|c_0 \ket = |000 \ket, |c_1 \ket = |111 \ket$
is generally regarded as able to correct all single bit flips, but no other errors.
However, one could instead use it to correct {\em all} double bit flips, at the expense
of the ability to correct {\em any} single bit errors.  The theorem above says that it
can do even more --- it can correct all two-bit errors of the same type.  Although this
might  seem surprising at first, it is easy to understand why it is true.  For this
code
$Z_r Z_s |c_j \ket = |c_j\ket$ so that $Z_rZ_s$ is degenerate with the identity.
Similarly, $Y_rY_s$ is degenerate with $X_rX_s$.   Note that this degeneracy
extends to any n-bit repetition code  
\be  \label{eq:rep}
|c_0\ket = |00 \ldots 0 \ket = W_0, \quad \quad |c_1\ket = |11 \ldots 1 \ket = W_n .
\ee

When $n \geq 5$, the simple repetition code (\ref{eq:rep})
can correct  all single and all double bit flips. Indeed, 
for $n = 5$, this is
just a classical code for two-bit error correction.
Applying the Hadamard code map (\ref{eq:Had}) to (\ref{eq:rep})
yields a code which can correct all single and double phase errors.  
In fact, omitting  the normalizing coefficients, this code is
\bsq  \label{eq:rep.phase} \be
  |C_0 \ket ~ = & H^{\ot n}|c_0 \ket +  H^{\ot n}|c_1 \ket
 & =  ~ \sum_{m} W_{2m} \\
  |C_1 \ket ~ = & H^{\ot n}|c_0 \ket - H^{\ot n}|c_1 \ket
 & = ~  \sum_{m} W_{2m+1}
\ee  \esq

Because the phase errors preserve parity, the necessary and sufficient conditions 
for a code satisfying conditions (I) and (II) to correct 
both single and double phase errors are 
\bsq \label{eq:2class} \be
0 &=& \bra c_0, \ovb{Z}c_0 \ket   \label{eq:2class.a}   \\
0 &=& \bra \ovb{Z}c_0, \ovb{ZZ}c_0 \ket \label{eq:2class.b}   \\
0 &=& \bra (Z_1-Z_t)c_0, Z_1Z_s c_0 \ket =0  \quad (t \neq s) . 
  \label{eq:2class.c}
 \ee  \esq
Note that since $\bra (Z_1-Z_t)c_0, \ovb{ZZ} c_0 \ket =0$, the
single and double-Z errors which transform as the $(n \mm 1)$-dimensional
representation are orthogonal if and only if (\ref{eq:2class.c}) holds.
In fact, as shown after (\ref{eq:AppZtZs}) in Appendix A, 
(\ref{eq:2class.c}) is redundant, i.e.,  it is satisfied
 whenever (\ref{eq:2class.a}) and (\ref{eq:2class.b}) hold.
Thus,  one finds that the
necessary and sufficient conditions for a code 
satisfying conditions (I) and (II) to correct single and double 
phase errors are (\ref{eq:skew}) [which becomes (\ref{eq:sum.im}) 
when $a_k$ is complex] and (\ref{eq: Z,ZZ}),
which we rewrite below.
\bsq  \label{seq:dbZ} \be
0 &= & \sum_k |a_k|^2 (n-2k)\binom{n}{k} =0 \label{eq:Z1Z2}\\
 0 &= & 
\sum_k |a_k|^2 (n-2k) [(n-2k)^2 - n] \binom{n}{k}=0 \label{eq: ZandZZ}.
\ee \esq

When $n = 5$, the pair of equations in (\ref{seq:dbZ}) has exactly
one solution (up to normalization), namely
$|a_0|^2 = |a_2|^2 =  |a_4|^2 $.  Choosing identical phases  yields
the code in (\ref{eq:rep.phase}).  In addition to correcting
all one and two-bit phase errors, it can also correct all
errors of the form $X_rX_s$ and $Y_r Y_s$.   Choosing other
phases yields other codes and taking the Hadamard code map
yields  classical codes for two-bit error correction that
are distinct from (\ref{eq:rep}).   These also
satisfy conditions (I) and (II) and, hence, can correct all 
double $Z_rZ_s$ and $Y_r Y_s$ errors as well as single and double bit 
flips when used as quantum codes.

When $n \geq 7$ and odd, the pair of equations (\ref{seq:dbZ})
has infinitely many solutions in addition to 
$|a_0|^2 = |a_2|^2 \ldots =  |a_{n-1}|^2 $.   Taking the
Hadamard transform then yields infinitely many classical
codes for two bit error correction.

\subsection{Higher dimensional representations}  \label{sect:higher}

In this section we take some preliminary steps toward exploiting higher 
dimensional irreducible representations for correction of errors {\em in 
addition} to one-bit errors.     
First, we review the mutually orthogonal subspaces
required for the correction of single errors.  
The operators $I, \ovb{X}, \ovb{Y}, \ovb{Z}$ acting on the code
words $|c_0 \ket, |c_1 \ket$ require four pairs of one-dimensional
subspaces which transform
as the trivial representation.  The three sets of differences 
$X_1-X_r$, $Y_1-Y_r$, 
and $Z_1-Z_r$ acting on the code words require
 three  pairs of  subspaces of dimension $n \mm 1$ which transform
as the even $(n \mm 1)$-dimensional representation.  But (as described in
Appendix B) the decomposition of $\bfC^{2^n}$ 
into an orthogonal sum of irreducible subspaces 
includes other irreducible  representations of $S_n$.

The next irreducible representation has dimension $\frac{n(n-3)}{2}$
and arises in the  decomposition  of double errors of one type,   e.g.,      
$f_{rs} = X_r X_s$.    Consider the subspace generated by 
$f_{rs}  W_k  $ for $k = 2$ (or $k= n - 2 $) as
$r,s$ run through all  $\frac{n(n-1)}{2}$  combinations of $r < s$.  
This  subspace splits into an orthogonal direct sum consisting of
\begin{itemize}
 \item a 1-dimensional subspace spanned by $\ovb{ff} \, W_k  $ where
$\ovb{ff} = \binom{n}{2}^{-1} \sum_{r \neq s} f_{rs}$ is the average error
of this type, 
 \item an $(n \mm 1)$-dimensional subspace spanned by the vectors $ f_r  W_k  $
 for $    r= 2,3 \ldots n$ where
$\ovb{f}_r = \sum_{s = 2}^n f_{1s} - \sum_{s \neq r} f_{rs}  ~~~~ r = 2, 3 \ldots n$ ,
   and
\item an $\frac{n(n-3)}{2}$-dimensional subspace  
which is obtained by taking the orthogonal complement of the  vectors
$\ovb{ff}  W_k  $ and $f_r  W_k  $ in   span$\{ f_{rs}  W_k   \}$ .

\end{itemize}
As described in Section~\ref{sect:err}, the error set $\{ f_{rs} \}$ can
be  correspondingly decomposed into bases for   representations of $S_n$
with dimensions $1$, $n - 1$, and $\frac{n(n-3)}{2}$.

For an explicit example of the last type of error,
 consider $n = 4$.  Then ${\cal W}_2$ splits into three subspaces,
corresponding to irreducible representations of dimensions 1, 3 and 2.
  The last is spanned by the vectors:
\bee
  2 f_{12} - f_{13} - f_{14} - f_{23} - f_{24} + 2 f_{34}\\
    \; \;\; f_{12} + f_{13} - 2f_{14} - 2 f_{23} + f_{24} +  f_{34}
\eee
There is a sense in which these errors are rather delocalized, since
they act on all six pairs of qubits.   Although one could eliminate some pairs
by a different choice of basis vectors, one can {\em not}, e.g.,
eliminate all terms of the form  $f_{j4}$ involving the 4th qubit.
This delocalization is, unfortunately, the antithesis of what one
might want in certain situations, such as errors between nearest neighbors.

We  now focus on  $n = 7$ as an example and note that
 $\bfC^{2^7}$ can be decomposed into an orthogonal
sum of irreducible subspaces spanned by 
\begin{itemize}
\item 8 orthogonal bases for the trivial 1-dimensional representation,
\item 6 orthogonal bases for the even 6-dimensional  representation,
\item 4 orthogonal bases for a 14-dimensional  representation, 
\item 2 orthogonal bases for a second, inequivalent, 14-dimensional representation.
\end{itemize}
Thus, correcting the one-bit errors requires all of the available $1$ and $6$
dimensional representations.   However, the two types of $14$-dimensional
representations are available to correct two-bit errors.   One of these $14$-dimensional
representations has spin $\frac{3}{2}$ and is associated with the partition $[5,2]$.
This is  the  $\frac{n(n-3)}{2}$-dimensional representation
which arises in the decomposition of   $f_{rs}$ errors described above,
and can be used to correct the corresponding subclass of double errors,
There are three kinds of double errors, those  from  $Z_rZ_s$,   from
$X_rX_s$, and  from $Y_r Y_s$.  This would seem to require 
six orthogonal subspaces which transform as this 14-dimensional
representation; however, we have only four --- one each from
 $\mW_2, \mW_3, \mW_4, \mW_5$.
Nevertheless, Theorems \ref{thm:doub.par} and  \ref{thm:doub} imply
that all three types of double errors can be corrected.

This is indeed the case and is the result of degeneracy.
 For permutationally invariant codes,
\be  \label{eq:deg}
  (X_r X_s + Z_r Z_s + Y_r Y_s) |\psi \ket = |\psi \ket .
\ee
Thus, it suffices to correct any two of $X_r X_s$, $Z_r Z_s$, $Y_r Y_s$
to ensure that all three types of errors can be corrected, and
this requires only four 14-dimensional subspaces, exactly what one
has available when $n = 7$.   Thus, a 7-bit permutationally 
invariant code which can correct all one-bit errors can also correct  
all  errors in the 14-dimensional irreducible components of
the decompositions of $X_r X_s$, $Z_r Z_s$, and $Y_r Y_s$.
This implies that arbitrary double errors  would be corrected about $2/3$ of the time.
Similarly, a 9-bit code could correct them about $3/4$ of the time.
Unfortunately, the other $1/3$ (or 1/4) of the time, the procedure does not
simply fail to detect the error --- it incorrectly interprets a two-bit
error as a one-bit error and the attempted correction actually introduces
additional errors.  

We next consider the case $n = 9$.  Correcting the one-bit errors
   uses 8 of
the 10 available $1$-dimensional representations and
6 of the 8 available $8$-dimensional representations.
In addition to the six $27$-dimensional
representations, two  $1$-dimensional representations and
two  $8$-dimensional representations are
also potentially available  to correct some two-bit errors.   
Correcting one type of $f_{rs}$ double errors would
require a pair of $1$-dimensional, $8$-dimensional and
$27$-dimensional subspaces.   Based on  dimensional 
considerations, one might expect to correct one type of double error
completely using a 9-bit permutationally invariant code.
Unfortunately, as will be shown in Section~\ref{sect:9no}, 
this is not possible.

One can still ask what additional errors can be corrected  
 with permutationally invariant 9-bit codes.
The operators $I, \ovb{X}, \ovb{Y}, \ovb{Z}$ acting on the code
words $|c_0 \ket, |c_1 \ket$ generate an  8-dimensional space.
Taking the orthogonal complement in the 10-dimensional subspace
spanned by $\{W_0, W_1 \ldots W_9\}$ yields a two-dimensional
subspace.   There is a family of linear operators which
map  $ |c_0 \ket,  |c_1 \ket$ to a pair of orthogonal vectors
in this two-dimensional subspace.  Any member of this
family can be chosen as an additional correctable error.
Similarly, there will be a set of correctable  errors which transform as
the 8-dimensional representation and whose action on the code
words spans the orthogonal complement of the one-bit errors 
in span$\big(\oplus_{k=1}^8 \mU_k^8 \big)$.  Although
a procedure for obtaining these operators can be written down,
we have been unable to characterize them in a  useful way.


\section{Special cases}

\subsection{n = 5}

When $n = 5$, conditions (I) and (II) hold, and all $a_k$ are real,
the three necessary and sufficient conditions in Theorem~\ref{thm:main}
become
\bee
   a_2 a_4 = 0   \label{eq:5a}  \\
   a_0 a_4 + 3a_2^2 = 0 \label{eq:5b}  \\
    a_0^2 + 2 a_2^2 - 3 a_4^2 = 0  .\label{eq:5c}
\eee
It is easy to verify that these have no non-trivial solution.  This
is not surprising.  It is well-known that the 5-bit code for correcting
all one-bit quantum errors is essentially unique and is {\em not}
permutationally symmetric.

Nevertheless, there is still something to be learned by looking at
5-bit codes.  As discussed in section~\ref{sect:deg}, the simple repetition code
\be  \label{eq:doub5}
   |c_0 \ket = |00000 \ket, ~~~~ |c_1 \ket = |11111 \ket
\ee
corrects both all single and all double bit flips, and
\bsq \label{eq:doub5H} \be
  |C_0 \ket ~ = & H^{\ot 5}|c_0 \ket +  H^{\ot 5}|c_1 \ket
 & = ~  W_0 + W_2 + W_4 \\
  |C_1 \ket ~ = & H^{\ot 5}|c_0 \ket - H^{\ot 5}|c_1 \ket
 & = ~  W_5 + W_3 + W_1
\ee  \esq
corrects all single and double phase errors.  In fact, when $n=5$, equations  
(\ref{eq:Z1Z2}) and 
(\ref{eq: ZandZZ}) imply that the {\em only} 
codes satisfying conditions (I) and (II) are those
with $|a_0|^2=|a_2|^2 =|a_4|^2$.

Moreover, both codes can correct {\em all} double errors of the form
$X_jX_k$, $Y_jY_k$, $Z_jZ_k$.  To see this, note that
$Z_r Z_s  |\psi \ket = |\psi \ket$ on the span of (\ref{eq:doub5})
so that (\ref{eq:deg}) implies
$X_r X_s |\psi \ket = - Y_r Y_s |\psi \ket$, i.e., the pair 
$\{ Z_r Z_s, I \}$ is degenerate and this induces a degeneracy on
the pair $\{ X_r X_s,  Y_r Y_s \}$.

Thus, the 5-bit codes (\ref{eq:doub5}) and (\ref{eq:doub5H}) can 
each correct more types of
quantum errors than one might expect  from their classical distance properties.  
They are optimal for the correction of all one-bit and two-bit errors of
a particular type (phase or bit flip) and can not correct additional
one-bit errors.   Nevertheless they can correct additional types
of two-bit errors.

\subsection{n = 7}

When $n = 7$, conditions (I) and (II) hold, and all $a_k$ are real,
the three conditions in Theorem~\ref{thm:main} become
\bsq \be
  3 a_2 a_6 + 5 a_4^2 = 0   \label{eq:7a}  \\
   a_0 a_6 + 15 a_2 a_4 = 0 \label{eq:7b}  \\
  a_0^2 + 9 a_2^2 - 5a_4^2 -5 a_6^2 = 0  \label{eq:7c} .
\ee  \esq
It is not hard to see that $a_6 = 0$ implies all $a_k = 0$.
Therefore we can divide through by $a_6$ or, equivalently,
assume without loss of generality that $a_6 = 1$.  Then (\ref{eq:7a}) and
(\ref{eq:7b})
imply $a_2 = - \frac{5}{3}a_4^2$ and $a_0 = 25 a_4^3$.  Letting
$x = a_4^2$ and substituting in  (\ref{eq:7c}) yields
$$
  125 x^3 + 5 x^2 - x - 1 = 0,
$$
to which $x = \frac{1}{5}$ is the only real solution, giving
$$
a_0=\pm \sqrt{5} \quad a_2=-\frac{1}{3} \quad a_4 = \pm \frac{1}{\sqrt{5}}
\quad a_6=1.
$$
It is then straightforward to verify that both signs in the normalized
vector
\be  \label{eq:code7}
|c_0 \ket & = &\tfrac{1}{8} \Big[ \pm \sqrt{15} \, \wh{W}_0 - \sqrt{7} \,
\wh{W}_2
    \pm \sqrt{21} \, \wh{W}_4 + \sqrt{21} \, \wh{W}_6 \Big] 
\ee
yield acceptable codes.  This gives two distinct new codes when $n = 7$.

It is interesting --- and a good check --- to write the vectors
$ \ovb{X} |c_1 \ket, \ovb{Y} |c_1 \ket,  \ovb{Z} |c_0 \ket$
and see that together with $|c_0 \ket$ they form an orthogonal set.
Up to normalizing scalars, we have
\bee
\ovb{Z} |c_0 \ket & = & \sqrt{\tfrac{1}{7}} \Big[ \pm \sqrt{35} \, \wh{W}_0
-
   \sqrt{3} \, \wh{W}_2 \mp  \wh{W}_4 -5 \wh{W}_6 \Big] \\
\ovb{X} |c_1 \ket & = & \sqrt{\tfrac{1}{7}} \Big[\sqrt{7} \, \wh{W}_0 +
   \sqrt{3}(2 \pm \sqrt{5}) \, \wh{W}_2 + (\pm 4 -\sqrt{5}) \,  \wh{W}_4
+( \pm \sqrt{5} -2)
\, \wh{W}_6 \Big]
\\
\ovb{Y} |c_1 \ket & = & \sqrt{\tfrac{1}{7}} \Big[-\sqrt{7} \, \wh{W}_0 +
   \sqrt{3}(2 \mp \sqrt{5}) \, \wh{W}_2 + (\pm 4+\sqrt{5}) \,  \wh{W}_4 +(
\mp \sqrt{5} -2)
\, \wh{W}_6  \Big]
\eee

\rmk  One might ask if one can obtain additional permutationally
invariant 7-bit codes by allowing complex coefficients.
In that case  the following equations are necessary and sufficient.
\bsq \be
0&=& 10|a_4|^2 + 3(\ovb{a}_2a_6+a_2\ovb{a}_6)  \label{cx3a} \\
0&=& \ovb{a}_2a_6-a_2\ovb{a}_6 \label{cx3b} \\
0&=&(\ovb{a}_0a_6+a_0\ovb{a}_6) + 15(\ovb{a}_2a_4 + a_2\ovb{a}_4) \label{cx3c} \\
0&=&(a_0\ovb{a}_6-\ovb{a}_0a_6) + 5(a_2\ovb{a}_4-\ovb{a}_2a_4) \label{cx3d} \\
0&=&|a_0|^2 + 9 |a_2|^2 -5|a_4|^2 - 5|a_6|^2 \label{cx3e} \\
0&=& (\ovb{a}_0a_2-a_0\ovb{a}_2) + 10(\ovb{a}_2a_4-a_2\ovb{a}_4) + 
5(\ovb{a}_4a_6-a_4\ovb{a}_6) \label{cx3f} .
\ee  \esq
As in the real case, 
$a_6=0$ forces all coefficients to be zero, so we can assume 
 $a_6=1$.  Then (\ref{cx3b}) implies that $a_2$ is 
real and (\ref{cx3b}) that $a_2=-(5/3)|a_4|^2$.   However,
(\ref{cx3d})  and (\ref{cx3f}) yield a pair of linear
equations for $\im a_0$ and $\im{a_4}$ which have a non-zero
solution if and only if $a_2 = +1$ which is not consistent with
$a_2=-(5/3)|a_4|^2$.    Thus, there are no 
permutationally invariant 7-bit codes other than those in (\ref{eq:code7}).

\subsection{n = 9}  \label{sect:9}

When $n = 9$, conditions (I) and (II) hold, and all $a_k$ are real,
the three conditions in Theorem~\ref{thm:main} become
\bsq \be
     a_2 a_8 +  7 a_4 a_6 = 0   \label{eq:9a}  \\
  35 a_4^2+ a_0a_8 + 28 a_2 a_6   = 0 \label{eq:9b}  \\
  a_0^2 + 20 a_2^2 + 14 a_4^2 - 28 a_6^2 - 7a_8^2= 0 . \label{eq:9c}
\ee  \esq
We now show that these have
 have infinitely many solutions.  First, suppose $a_8=0$, so
that the first equation becomes $a_4a_6=0$.  If $a_6=0$, then all the coefficients are
zero. If $a_4=0$ then we find the only possibility is $a_2=0$ and $a_0^2=28a_6^2$,
giving the two solutions found in \cite{Paul1}.
To find the remaining solutions we may assume $a_8=1$.

If $a_6=0$, then also $a_2=0$ and $a_0=-35a_4^2$ where $a_4^2=x$ is the
positive root of the quadratic equation
$175x^2 +2x-1=0$.  This gives two more
solutions.
If $a_0=0$, then we find all of the remaining coefficients depend on $a_6^2=t$, where
$t$ is a  positive root of the cubic $f(t)=(28^3/5)t^3+(2\cdot 28^2/5)t^2-4t-1$:
$$
a_4=(28/5)t, \quad a_6=\pm\sqrt{t}, \quad a_2 = (- 7 \cdot 28)t.
$$
Since $f(0)<0$ and $f(1)>0$, $f$ does have a positive root (approximately $t=0.478$), and this gives
two more solutions.

There are no further solutions with any of the $a_k$ equal to zero, so now
assume they are all nonzero.  Writing $x=a_6^2$ and $t=a_4$, it follows that
$$
a_2= \pm 7 \sqrt{x}t \quad a_0=7(28tx-5t^2)
$$
where  $x$ and $t$ satisfy the equation
$$
x^2(5488t^2) + 4x(-490t^3 + 35t^2-1) + (175t^4+2t^2-1)=0.
$$
Using Maple, one can verify that there are infinitely many values of $t$
(e.g., all for which $-0.25<t<0.4$) for which this quadratic in $x$ has at 
least one positive solution.  
Thus there are infinitely many solutions in which all the coefficients 
$a_{2m}$ are non-zero real numbers.

\subsection{Conditions for double error correction with 9-bit codes} 

As discussed in
Section~\ref{sect:twobt.cond}, a 9-bit permutationally 
invariant codes which can correct all errors of type $Z_r Z_s$ as
well as all one-bit errors, must satisfy   
 at least 9 conditions.  In the notation of Section~\ref{sect:basic}, 
there are  six of the form
$ b_{IX} = b_{IY} = b_{ZX} = b_{ZY} = b_{ZZ,X} = b_{ZZ,Y} = 0$ and
three of the form $ d_{XY} = d_{IZ} =    d_{Z,ZZ} = 0 $.

First, consider the  conditions (\ref{sum1a}), 
(\ref{sum2a}) and  (\ref{XYp.ZZ})
which correspond to the requirements 
$\bra \ovb{f} c_0 , ( \ovb{X} + i \ovb{Y}) \, c_1 \ket = 0$
with $f = I, Z$ or $ZZ$.  For $n = 9 $ these are equivalent to
\bsq \label{seq:dbp} \be  \label{eq:dbpa}
  \ovb{a}_2  a_8 + 7 \ovb{a}_4 a_6 & + 7 \ovb{a}_6 a_4 & + \ovb{a}_8 a_2 = 0 \\
5 \ovb{a}_2  a_8 + 7 \ovb{a}_4 a_6 & - 21 \ovb{a}_6 a_4 &   - 7 \ovb{a}_8 a_2 = 0
\label{eq:dbpb} \\ 
  2 \ovb{a}_2  a_8 - 7 \ovb{a}_4 a_6 & ~ &  +5 \ovb{a}_8 a_2 = 0  .
  \label{eq:dbpc}
\ee  \esq
These can be treated as a set of 3 linear equations in the 4 unknowns,
$\ovb{a}_2  a_8$, $ \ovb{a}_4 a_6, \ovb{a}_6 a_4,  \ovb{a}_8 a_2$ from
which  one finds that the group (\ref{seq:dbp}) is equivalent to
\bsq  \label{seq:dbpnua} \be 
    \ovb{a}_2  a_8 = -  \ovb{a}_8 a_2 = -  i \nu   \label{eq:dbpnua_1}\\
    \ovb{a}_4 a_6 = -   \ovb{a}_6 a_4 =  i \frac{3}{7} \, \nu  
\label{eq:dbpnua}
\ee   \esq
for some real parameter $\nu$.   
Note also that $\re \ovb{a}_4 a_6  = \re  \ovb{a}_2  a_8 = 0$ implies
that any real solutions must have $a_4 a_6 = a_2 a_8 = 0$. 
However, all such solutions have been found above and none  
satisfy  the additional requirements below. 
   Hence, correcting all double-$Z$ errors does require complex
coefficients. 

\medskip

Next we consider  
the conditions (\ref{diff1a}), (\ref{diff2a}) and  (\ref{XYm.ZZ})
which correspond to the requirements 
$\bra \ovb{f}  c_0 , ( \ovb{X} - i \ovb{Y}) \, c_1 \ket = 0$
with $f = I, Z$ or $ZZ$.  These become
\bsq  \label{seq:dbm} \be
  \ovb{a}_0  a_8 + 28 \ovb{a}_2 a_6   + 70|a_4|^2 & + 28 \ovb{a}_6 a_2 & + \ovb{a}_8 a_0
   = 0    \label{eq:ZZ2a} \\ 
9 \ovb{a}_0  a_8 + 140 \ovb{a}_2 a_6   + 70|a_4|^2 & - 84 \ovb{a}_6 a_2 & -7 \ovb{a}_8 a_0  
   = 0    \label{eq:ZZ2b} \\ 
9  \ovb{a}_0  a_8 + 56  \ovb{a}_2 a_6   - 70|a_4|^2 &   & + 5 \ovb{a}_8 a_0 
 = 0    \label{eq:ZZ2c}  
\ee  \esq
which can be rewritten as
\bsq \label{seq:dbm.alt}  \be
  \re \ovb{a}_0  a_8 + 28 \re \ovb{a}_2 a_6  + & 35 |a_4|^2 &  = 0 
    \label{seq:dbm.alt.a} \\
  7 \re \ovb{a}_0  a_8 + 28 \re \ovb{a}_2 a_6  - & 35  |a_4|^2 &  = 0 
  \label{seq:dbm.alt.b} \\
  \im \ovb{a}_0  a_8 + 14 \im \ovb{a}_2 a_6 && = 0   \label{seq:dbm.alt.c}
\ee  \esq
since (\ref{eq:ZZ2a}) and (\ref{eq:ZZ2b}) have the same real part
while (\ref{eq:ZZ2b}) and (\ref{eq:ZZ2c}) have the same imaginary part.
Equations (\ref{seq:dbm.alt.a}) and (\ref{seq:dbm.alt.b}) are equivalent
to
\bsq \label{eqa6alt} \be
\re\ovb{a}_0a_8&=&\frac{35}{3} |a_4|^2 \\
\re\ovb{a}_2a_6& = &-\frac{5}{3}|a_4|^2 .
\ee  \esq
 
To these conditions we need to add the requirements
$ \bra \ovb{X}c_j , i \, \ovb{Y} c_j \ket =
  \bra c_j ,  \ovb{Z}c_j \ket  = \bra \ovb{Z} c_j ,  \ovb{ZZ}c_j \ket =
0$ which become.  
\bsq  \label{seq:dZZ} \be   \label{eq:dZZa}
 |a_0|^2 + 20 |a_2|^2 + 14 |a_4|^2   - 28 |a_6|^2   -  7|a_8|^2  & = &
0   \\ \label{eq:dZZb}
  9 |a_0|^2 + 40 |a_2|^2 - 14 |a_4|^2 \hskip1.5cm -  35  |a_8|^2   & = & 0
    \label{eq:dbIZZ}  \\    \label{eq:ImXY}
  \im \ovb{a}_0 a_2 + 21 \im \ovb{a}_2 a_4 + 35 \im \ovb{a}_4 a_6 +
7 \im \ovb{a}_6 a_8 & = & 0.
\ee  \esq
The last equation (\ref{eq:ImXY}) is precisely the condition
$\im \bra \ovb{X}c_j , i \, \ovb{Y} c_j \ket = 0$ 
and is obtained as the reduction
of (\ref{eq:XYcomp}) when $n = 9$ and conditions  I and II hold.  

To recap, we have three groups of equations; namely, (\ref{seq:dbp}) 
from the conditions $b_{fX}+ i b_{fY} = 0$, (\ref{seq:dbm}) from the conditions
$b_{fX} - i b_{fY} = 0$, and  (\ref{seq:dZZ}) from the conditions
$ d_{XY} = d_{IZ} =    d_{Z,ZZ} = 0 $.   In what follows, we will use
the equivalent conditions  (\ref{seq:dbpnua}) in place of (\ref{seq:dbp}),
and (\ref{eqa6alt})   or (\ref{seq:dbm.alt.c})
in place of (\ref{seq:dbm}).

\subsection{Limits on correction of $\ovb{ZZ}$ and one-bit errors}  \label{sect:9no}

To analyze the conditions obtained above, write $a_k = x_k + i y_k$.  
We can assume without loss of generality that $a_8 = 1$; then 
(\ref{eq:dbpnua_1}) implies that $a_2 = i \nu$ 
and (\ref{eqa6alt}) implies  $y_6 = - \frac{5}{3 \nu} |a_4|^2$.
We also have that (\ref{eq:dbpnua}) implies
$a_6 =\frac{3\nu}{7} \frac{1}{|a_4|^2} a_4 i $
from which we can conclude
$y_6 =\frac{3\nu}{7} \frac{x_4}{|a_4|^2} $.
Equating the two expressions above for $y_6$  yields
\be
\nu^2 &=& - \frac{35}{9}\frac{|a_4|^4}{x_4} \label{nueq} 
\ee
which implies that $x_4<0$. Under the assumption $a_8 = 1$, 
 (\ref{eq:ImXY}) becomes
\be  \nn 
0 &=& \nu \, \Big( x_0  - 21  x_4  + 15   - \frac{7}{\nu} y_6 \Big) \\
 &=& \nu \, \Big(\frac{35}{3}|a_4|^2 +     21 |x_4| + 15 +
     \frac{35}{3 \nu^2} |a_4|^2 \Big)
 \label{eq6}
\ee
The last equation implies that either $\nu = 0$ or $a_4=0$, either of 
which generates  only a trivial solution.  Thus there is {\em no
non-trivial solution} to the seven equations 
 (\ref{seq:dbp}),  (\ref{seq:dbm}) and (\ref{eq:ImXY}).

By Theorem~\ref{thm:XZequiv}, this implies that there is no 9-bit 
permutationally invariant
code which can correct all one-bit errors as well as one type of
double error.    

One might wonder if there is a 9-bit code which satisfies all
the conditions above, except (\ref{eq:ImXY}).  Such a code would
still be of some interest.  It would be able to correct all
single  and double $Z$ errors, and detect all single $X$ and 
$Y$ errors.  However, it would not be able to correct
$X_k$ and  $Y_k$ errors because it could not reliably distinguish
between them.   Unfortunately, even this is not possible.

We return to
the equations (\ref{seq:dbp})  and (\ref{seq:dbm}) and observe that
there are infinitely many solutions that can be expressed using
one complex variable $a_4$,  or two real variables $x_4, y_4$, in 
either case with the constraint  $\im a_4 = x_4 < 0$.   
Let $x = - x_4 > 0$ and $y = y_4$.  Then we have
\begin{align*}
  a_0 & =   \frac{35}{3} \big(1 - 2 i \frac{y_4}{x_4} \big) |a_4|^2 
  &  |a_0|^2 & =    
    \frac{35^2}{9} \frac{x^2 + 4y^2}{x^2}(x^2 + y^2)^2\\
  a_2 & =  i \nu  = \pm i \frac{\sqrt{35}}{3} \frac{|a_4|^2}{\sqrt{|x_4|}} 
     &  |a_2|^2 & = 
   \frac{35}{9} \frac{(x^2 + y^2)^2}{x}  \\
   a_4 & =    x_4 + iy_4   & |a_4|^2  &  =  x^2 + y^2 \\
   a_6 & =   \pm i \frac{\sqrt{35}}{7} \frac{a_4}{\sqrt{|x_4|}} 
      & |a_6|^2 & = 
   \frac{5}{7}   \frac{x^2 + y^2}{x}  \\
  a_8 & = 1  & |a_8|^2 & = 1 .
\end{align*}
Substituting into (\ref{seq:dZZ}a) and (\ref{eq:dbIZZ}) yields
 two equations in two unknowns which have no solution.
Thus, there is no 9-bit code which satisfies all the 
desired equations except 
(\ref{eq:ImXY}).

We also considered the possibility of dropping all $Y_k$
conditions  to find a code which could correct all errors
of the form  single $X_k$,  single $Z_k$ and double $Z_j Z_k$.
However, this is as restrictive as dropping only (\ref{eq:ImXY}).

\section{Concluding Remarks}

Permutationally invariant codes which can correct all one-bit
errors require a minimum of seven qubits.  We have shown that there
are two distinct 7-bit codes of this type.     Although one might expect
that 9-bit codes could also correct one class of double errors,
a detailed analysis shows that this is not possible.
Even a 9-bit code which could correct  all 
one-bit errors of the form $X_k$ and $Z_k$  and all two
bit errors of the form $Z_j Z_k$ does not  exist.
If one modifies this to the requirement that the code
be able to correct all one bit errors of the form $X_k$
and double errors of the form $X_j X_k$ and $Z_j Z_k$,
this can be done.  However, it does {\em not} require 9-bits;
it can be achieved using the simple 5-bit repetition code
(\ref{eq:doub5}) which can correct all double errors of the form 
$X_j X_k$ and $Y_j Y_k$ as well. 

Permutationally invariant codes are highly degenerate, since
all $\binom{n}{2}$ exchange errors are equivalent to the identity.
As discussed in Section~\ref{sect:deg}, and illustrated by the
5-bit repetition code, this degeneracy can sometimes lead to enhanced 
ability to correct two-bit errors.  However, there are also
limitations on their ability to correct all two-bit errors of a
given type as well as all one-bit errors, as shown by our analysis of
of 9-bit codes.   Although the reasons for this remain unclear, it 
may be that the ``degeneracy enhancement'' also gives hidden
constraints, i.e.,  that one is implicitly trying to correct
more two-bit errors than those from which the conditions 
were obtained.   

We have concentrated here on the construction of permutationally
invariant codes.   Actual implementation would require a 
number of additional considerations.  For example, one would
need a mechanism for initializing the computer in states
corresponding to $\kt{c_0} \ot \kt{c_0} \ldots \kt{c_0}$.
One could then obtain any state of the form
$\kt{c_{k_1}} \ot \kt{c_{k_2}} \ldots \kt{c_{k_m}}$ with
$k_i \in \{0,1\}$ by application of 
 $\ot_{j = 1}^{n} X_{\kappa n + j}$ for suitable 
choices of $\kappa$.  
One also needs a mechanism for decoding, including a set of
measurements which can distinguish  between the different
error subspaces,  as well as a circuit for implementing
the error correction process.

Finally, one needs a set of gates for universal computation.
  As noted in Section~\ref{sect:code},
the logical  $X$ and $Z$ operations are easily implemented
as products of their single-bit counterparts.   The logical
$Y$ is given by the product
$ i [\ot_j X_j][\ot_j Z_j]  =  (-1)^{(n-1)/2} [\ot_j Y_j]$ 
when $n$ is odd.   Moreover, the
$X,Y,Z$ gates all lie in the commutant of $S_n$.  This implies
that all gates needed for universal computation (including a
non-trivial two-bit gate) also lie in the commutant since they
can be written  as linear combinations of these operations, the
identity, and their products.  (Alternatively, one could observe
that all logical actions  on code words must be permutationally 
symmetric and, hence, lie in the commutant of $S_n$.)
  However, this does not necessarily mean that all the desired
gates can be implemented as products of a small set of one and 
two-bit gates.   We leave the question of a practical implementation
of a universal set of gates on code words for further investigation.

We have only begun to explore the potential of non-Abelian stabilizer 
codes for quantum error correction; other examples
should be studied.  In addition to the issues
identified above, there may be others which arise if one wants to combine 
non-Abelian stabilizers with other approaches to fault tolerant 
computation.

\bigskip 

 \pagebreak

\noindent{\bf Acknowledgment:}  Parts of this work were done
while MBR was an exchange professor at the University of Massachusetts
Amherst and while HP was at the University of Sussex and 
Queen Mary College of the University of London during her sabbatical.
The authors are grateful to these institutions for their hospitality.

\bigskip 


\appendix

\section{Differences of one-bit errors}  \label{sect:appA}

In this section we will need some additional notation.  
Let $  \vep_r  $ denote the binary n-tuple  
  with components $\vep_j = \delta_{jr}$ 
so that $ v+ \vep_r $ has components
$v_j + \delta_{jr}$ with addition  mod 2.  Let
$ \ovb{1}  $ be the  binary n-tuple   with all elements equal to $1$.
We will use
$\sp(v) =  \{ j \, : \, v_j \neq 0 \}$ to denote the support of 
 of $v=(v_1, \ldots, v_n)$.

It will be convenient to also introduce the vector
\be    \label{eq:Vm.def}
V_{k}(r,s)  = \sum_{\sa{\wt(v) = k}{r, s \notin \sp(v)}} (|v+\vep_r\ket -|v +\vep_s
\ket)
\ee
which is well-defined for $k = 0, 1, \ldots (n-2)$, and has the following properties when $r \neq s$.
\be
\bra V_k(r,s)  ,  W_{\ell}\ket  &=& 0 ~ ~ \hbox{for all} ~ r,s,k,\ell \\
\bra V_{k}(r,s) ,  V_{\ell}(q,t) \ket &=& 0 ~ ~\hbox{for} ~ m \neq \ell ~
\forall ~
r,s,q,t \\
\bra V_{k}(r,s) ,  V_{k}(r,s) \ket &=&  2 \,  \binom{n-2}{k}
\label{eq:Vrs} \\
\bra V_{k}(r,s) ,  V_{k}(r,t) \ket &=&   \binom{n-2}{k}
~ \hbox{for} ~ r,s,t ~\hbox{all distinct} .  \label{eq:Vrst}
\ee
These are all straightforward, except (\ref{eq:Vrst}) which follows from
\bee
\bra V_{k}(r,s) ,  V_{k}(r,t) \ket =
 \binom{n-3}{k} + \binom{n-3}{k-1} \label{XYrst}
\eee
and the easily verified combinatoric identity
$\binom{n-2}{k} = \binom{n-3}{k} + \binom{n-3}{k-1} $.

An important consequence of   (\ref{eq:Vrs}) and (\ref{eq:Vrst})
is that they imply that, for $s \neq t$,  
\be  \label{eq:fact2}
\bra V_{k}(r,s) ,  V_{k}(r,s) \ket = 2 \, \bra V_{k}(r,s) ,  V_{k}(r,t) \ket.
\ee
This result plays an essential role in section~\ref{sect:skew.blk}.

Our main results are that, for any code of the general form (\ref{eq:code.gen}),
\bsq \be  \label{eq:ZVk}
(Z_r - Z_s) |c_0 \ket & = &-2  \sum_{k=1}^{n-1} a_k \, V_{k-1}(r,s)  \\
 (X_r - X_s)|c_0 \ket &=& \sum_{k =0}^{n-2 } (a_{k} - a_{k + 2})
 \,  V_{k}(r,s)    \label{eq:XVk} \\
   i(Y_r - Y_s)|c_0 \ket &=& \sum_{k=0}^{n-2}
   (a_{k} + a_{k + 2}) \, V_{k}(r,s)    \label{eq:YVk}
\ee  \esq
with similar equations for $|c_1 \ket$ and $b_k$.
Under the assumption that
conditions (I) and (II) hold, we find the following variants useful
\bsq    \label{eq:ZVm} \be
(Z_r - Z_s) |c_1 \ket &=&  -2 \sum_{m=0}^{(n-1)/2} a_{n-2m-1} \,
 V_{2m}(r,s)  \label{eq:dfZm} \\
 \big[ (X_r - X_s) + i \,(Y_r - Y_s)\big] \, |c_0 \ket  & = &
  ~ 2 \sum_{m=0}^{(n-3)/2}  a_{2m}  \,   V_{2m}(r,s)  \label{eq:dfXYp} \\
  \big[ (X_r - X_s) - i \,(Y_r - Y_s)\big] \, |c_0 \ket  & = &
  - 2 \sum_{m=1 }^{(n-1)/2}  a_{2m}  \,   V_{2m-2}(r,s) \label{eq:dfXYm} .
\ee  \esq

To prove (\ref{eq:ZVk}) and (\ref{eq:ZVm}) in the case of
$Z_r$, it
suffices to observe that
\be
(Z_r-Z _s)W_k = \left\{ \begin{array}{lcl} 0  & \mbox{for} & k = 0,n   \\
  -2V_{k-1}(r,s) & \mbox{for} &  1 \leq k \leq n-1  \end{array} \right.
\ee
which is easily verified.

Equations (\ref{eq:XVk}) and (\ref{eq:YVk}) can be verified by some rather
straightforward, but tedious,
computations and combinatorics.   One approach
is to write
out the effect of the errors $X_r$ and $Y_r$.  Since these results are
identical
except for the signs of some terms, we introduce
\be
\omega_{XY} \equiv \left\{ \begin{array}{ccc} +1 & \hbox{for} & X \\
      -1 & \hbox{for} & i Y  \end{array} \right.
\ee
and write the equations only for $X_r$ with the understanding that
these results hold for $Y_r$ with the sign changes indicated by
$\omega_{XY} $
\be
X_r W_0 &=& |\vep_r \ket \nn \\
X_r W_k &=& \sum_{\sa{\wt(v)=k}{ r \notin \sp(v)}} | v + \vep_r \ket ~ + ~
\omega_{XY}  \!\!
\sum_{\sa{\wt(u)=k-1}{r \notin s(u)}} |u \ket \quad \mbox{for  }
 \; 1 \leq k \leq n-1 \nn
\\ X_r W_n &=& \omega_{XY} | \ovb{1} + \vep_r \ket. \label{Xrfirst} \nn
\ee
For distinct $r$, $s$, we want to determine the effect of the differences
$X_r - X_s$, $Y_r - Y_s$  on the $W_k$, and for this purpose, the
following expression, which we write only for $2 \leq k \leq n-2$,
is useful.
\be
X_r W_k &=& \sum_{\sa{\wt(v)=k}{r, s \notin \sp(v)}}\!\! |v + \vep_r \ket 
 + \sum_{\sa{\wt(v) = k-1}{r, s \notin \sp(v)}} \!\! |v + \vep_s + \vep_r \ket \nn
\\ & ~ &
 + ~ \omega_{XY} \!\! \sum_{\sa{\wt(u)=k-1}{r, s \notin s(u)}} \!\! |u \ket + ~
 \omega_{XY}  \!\! \sum_{\sa{\wt(u)=k-2}{r, s \notin s(u)}} \!\! |u + \vep_s \ket
\label{Xr}
\ee
Then (\ref{Xr}) implies
\be
(X_r - X_s) W_0 &=& |\vep_r \ket - |\vep_s \ket   = V_0(r,s) \nn \\
(X_r - X_s) W_1 &=& \sum_{j \neq r, s} (|\vep_j + \vep_r\ket - |\vep_j + \vep_s \ket)
  = V_1(r,s) \nn \\
(X_r - X_s) W_k &=&  V_k(r,s) - \omega_{XY}  V_{k-2}(r,s) \nn \\
(X_r - X_s) W_{n-1}&=& \omega_{XY}
  \sum_{j \neq r, s} (|\ovb{1} + \vep_j + \vep_r\ket - |\ovb{1} + \vep_j + \vep_s \ket)
  = - \omega_{XY} V_{n-3}(r,s)
\nn \\
(X_r - X_s) W_n &=& \omega_{XY} (| \ovb{1} + \vep_r \ket - | \ovb{1} +
\vep_s \ket )  = - \omega_{XY} V_{n-2}(r,s) .
  \nn \label{Xrs}
\ee

 \medskip

To analyze double phase errors,  first
observe that when $\bra  \ovb{f} c_i, \ovb{ZZ} c_j \ket = 0$
(with  $f = X, Y, Z$),
the analogous inner products involving $(n \mm 1)$-dimensional
representations will be zero if and only if
$\bra (f_1-f_t)c_i, Z_r Z_s c_j \ket = 0$.
By considering the action of the transposition $(1t)$, one
can show that this holds  whenever
  $\{1,t\}=\{r,s\}$  or   
$\{1,t\} \cap \{r,s\} = \emptyset$.
Hence, it suffices to consider $r = 1$ and $t \neq s$, in which case one
 can use (\ref{eq:ZVk}) and $Z_1Z_s = I - Z_1(Z_1-Z_s)$ to conclude that
\be
Z_1Z_s |c_0 \ket = |c_0 \ket + 2 \sum_{k=1}^{n-1} a_k Z_1  V_{k-1}(1,s).
\ee
We will also need the formula
\be  \label{eq:Z1VV}
  \bra Z_1V_k(1,s), V_k(1,t) \ket = 
  \binom{n-3}{k-1}-\binom{n-3}{k}  
  =  \frac{2k+2-n}{n-2} \binom{n-2}{k} 
\ee
 which follows from (\ref{eq:comb}) with $N = n-3$.

We now let $f = Z$. 
 Using (\ref{eq:ZVk}) again and (\ref{eq:Z1VV}) with $k = 2m-1$,
one finds  
 $\bra (Z_1-Z_t)c_0, Z_1 Z_s c_0 \ket = 0$
if and only if
\be  \label{eq:AppZtZs} 
\sum_{m=1}^{(n-1)/2} |a_{2m}|^2 \frac{4m - n}{n-2} \binom{n-2}{2m} = 0 . 
\ee
Then it follows from 
$ \frac{2k - n}{n - 2} \binom{n-2}{k-1} 
  =  \frac{(2k-n)k(n-k)}{n(n-1)(n-2)} \, \binom{n}{k}$
that  (\ref{eq:AppZtZs}) is equivalent to  
(\ref{eq:2class.b})  minus $n^2 \mm n$ times equation (\ref{eq:2class.a}).   
Thus, the ``block'' conditions for double phase errors do not add
additional constraints when conditions (I) and (II) hold.

The cases $f = X \pm i Y$, and $i = 0, j = 1$, can be dealt
with similarly, but are not needed here. We note only that,
unlike the case $f = Z$, they do generate additional constraints. 



\section{Decomposition into irreducibles}  \label{sect:irred}

In view of the repeated use of decompositions into irreducible
subspaces, we explicitly write out some of them.
Recall that $\bfC^{2^n} = \oplus_{k = 0}^n \, \mW_k$; each
$\mW_k$ is the eigenspace of the operator  
$S_z  \equiv  \half \sum_k Z_k = \frac{n}{2} \ovb{Z}$ with eigenvalue $\half(n - 2k)$.
Each 
$\mW_k$ can be further decomposed into a direct sum of subspaces
which transform as irreducible representations of $S_n$.    In fact, 
the action of $S_n$ on $\mW_k$ or $\mW_{n-k}$ is its action on sets of
size $k$ in 
$\{1, \ldots, n\}$.  For $0 \leq k \leq \lfloor \frac{n}{2}\rfloor$, $\mW_k$ 
is known to decompose into a sum of irreducible subspaces indexed by the 
partitions $[n-j, j]$ for $j=0, \ldots, k$, 
each appearing once.   Physicists may recognize that this is 
equivalent to a decomposition into simultaneous
eigenspaces of $S_z$ and the total spin\footnote{In
this one paragraph, we use the familiar  $ S_x, S_y, S_z$,
rather than the equivalent   $\frac{n}{2} \ovb{X},  \frac{n}{2} \ovb{Y},  \frac{n}{2} \ovb{Z}$,
to denote the components of spin,    and trust that context suffices to distinguish
them from  the symmetric group denoted $S_n$, which is a very
different entity. }   operator
  ${\bf S}^2 = S_x^2 + S_y^2 + S_z^2$ with eigenvalue $s(s+1)$ for 
$s = \frac{n}{2},  \frac{n}{2} - 1,  \ldots \half |n - 2k|$.
 
 \smallskip

 To facilitate use in counting arguments, as in Section~\ref{sect:higher}, we 
adopt the convention of labeling an irreducible subspace (in part) by its
dimension.   Thus,  
 $\mU_k^d$ denotes an irreducible subspace of $\mW_k$
 with dimension  $d$.  

\medskip

 For $n = 5$
\bee
  \mW_0 & = & \mU_0^1 \\
   \mW_1 & = & \mU_1^1 \oplus \mU_1^4 \\
  \mW_2 & = & \mU_2^1 \oplus \mU_2^4 \oplus \mU_2^5 \\
  \mW_3 & = & \mU_3^1 \oplus \mU_3^4 \oplus \mU_3^5 \\
  \mW_4 & = & \mU_4^1 \oplus \mU_4^4 \\
  \mW_5 & = & \mU_5^1 
\eee

\medskip

 For $n = 7$
\bee
  \mW_0 & = & \mU_0^1 \\
   \mW_1 & = & \mU_1^1 \oplus \mU_1^6 \\
  \mW_2 & = & \mU_2^1 \oplus \mU_2^6 \oplus \mU_2^{14}\\
 \mW_3 & = & \mU_3^1 \oplus \mU_3^6 \oplus \mU_3^{14}  \oplus \mU_4^{\wtd{14}} \\
\mW_4 & = & \mU_4^1 \oplus \mU_4^6 \oplus \mU_4^{14}  \oplus \mU_4^{\wtd{14}} \\
  \mW_5 & = & \mU_5^1 \oplus \mU_5^6 \oplus \mU_5^{14} \\
  \mW_6 & = & \mU_6^1 \oplus \mU_6^6 \\
  \mW_7 & = & \mU_7^1 
\eee
where $\mU_2^{14}$ denotes the irreducible representation associated 
with the partition $[5,2]$ and  $\mU_4^{\wtd{14}}$ 
 a second, distinct, 14-dimensional irreducible
representation associated with
$[4,3]$.  In terms of spin, $\mU_2^{14}$ has $s = \frac{3}{2}$ and
$\mU_4^{\wtd{14}}$ has $s = \half$.
\medskip

 For $n = 9$
\bee
  \mW_0 & = & \mU_0^1 \\
   \mW_1 & = & \mU_1^1 \oplus \mU_1^8 \\
  \mW_2 & = & \mU_2^1 \oplus \mU_2^8 \oplus \mU_2^{27}\\
 \mW_3 & = & \mU_3^1 \oplus \mU_3^8 \oplus \mU_3^{27}  \oplus \mU_4^{48} \\
\mW_4 & = & \mU_4^1 \oplus \mU_4^8 \oplus \mU_4^{27}  \oplus \mU_4^{48} \oplus \mU_4^{42} 
\\
\mW_5 & = & \mU_5^1 \oplus \mU_5^8 \oplus \mU_5^{27}  \oplus \mU_5^{48} \oplus
\mU_4^{42} 
\\
\mW_6 & = & \mU_6^1 \oplus \mU_6^8 \oplus \mU_6^{27}  \oplus \mU_4^{48} \\
  \mW_7 & = & \mU_7^1 \oplus \mU_7^8 \oplus \mU_7^{27} 
\\
  \mW_8 & = & \mU_8^1 \oplus \mU_8^8 
\\  \mW_9 & = & \mU_9^1 
\eee

\smallskip


\section{Complex coefficients} \label{app:compx}

If the $a_k$ are not real, then one must modify the analysis
in Section~\ref{sect:anal} accordingly, and require
both real and imaginary parts of the resulting equations
to be zero.
 We again  use the classification of error conditions 
described at the end of Section~\ref{sect:basic}.   
 We omit the details and summarize the results.

\begin{itemize}

\item[a)]  Setting the real parts of  (\ref{sum1a})-(\ref{diff2a}) to zero
yields the pair of equations
\be  \label{eq:sum.re}
  0 & = & {\scriptsize \frac{n+1}{2}}   \binom{n}{\frac{n+1}{2}}
 a_{\frac{n+1}{2}}^2
 ~ +  2 \sum_{k = 1}^{(n-1)/2}   k \binom{n}{k} \, 
{\rm Re}( \ovb{a}_k a_{n-k+1})  \\ 
0 & = & {\scriptsize \frac{n+1}{2}}  
\binom{n}{\frac{n-1}{2}} a_{\frac{n-1}{2}}^2
 ~ + 2 \sum_{k = 0}^{(n-3)/2}    (n-k)
\binom{n}{k} \, {\rm Re}(\ovb{a}_k a_{n-k-1} ) . ~~~~~
\ee
The imaginary parts of both  (\ref{sum1a}) and  (\ref{diff1a}) are
always zero and do not place any additional restrictions on $a_k$.
Setting the imaginary parts of  (\ref{sum2a}) and (\ref{diff2a})
to zero yields the conditions
 \be \label{eq:sum.im}
 0 & = & \sum_{k = 1}^{(n-1)/2} {\rm Im}[ \ovb{a}_k a_{n-k+1}]
    \,  k(n \! - \!2k\!+\!1) \, \binom{n}{k} \\  \label{eq:diff.im}
  0 & = & \sum_{k = 0}^{(n-1)/2} {\rm Im}[ \ovb{a}_k a_{n-k-1}]
    \,  (n-k)(n \! - \!2k\!-\!1) \, \binom{n}{k}.
\ee

\item[b)] The condition (\ref{eq:skew})
from the off diagonal terms in $D$ becomes
\be  \label{eq:IZcomp}
 0 = \sum_{k = 0}^{n} |a_{k}|^2 (n-2k) \binom{n}{k} .
\ee
(This   is sufficient to ensure that  $ d_{IZ} =0$ 
and  $D_{IZ} = 0$, as well as that the real part of (\ref{eq:diagss})  
is zero.)  To ensure that the imaginary part of 
$\bra \ovb{X}c_i ,  i \, \ovb{Y} c_i \ket$ is zero we must also require
\be   \label{eq:XYcomp}
 0 & = & \sum_{k = 1}^{n-1} {\rm Im}[ \ovb{a}_{k+1} a_{k-1}]. 
\binom{n-2}{k-1} .
\ee
(This also ensures that the imaginary part of (\ref{eq:diagss}) is
zero so that  $ d_{XY} =0$  and  $D_{XY} = 0$.)

\end{itemize}
As before, we analyze the ``block'' conditions only under
the assumption that conditions (I) and (II) hold.   As for
real coefficients, these conditions do not yield new requirements.
\begin{itemize}

\item[c)] Setting the imaginary parts of (\ref{eq:XYblk1}) and
(\ref{eq:XYblk2}) to zero yields conditions equivalent to
(\ref{eq:sum.im}) and (\ref{eq:diff.im}).

\item[d)] The expression in (\ref{eq:diagss}) gives two conditions.    
The first is equivalent to (\ref{eq:IZcomp}) and 
the second to (\ref{eq:XYcomp}) with $k = 2m+1$.

\end{itemize}

  \bigskip


\end{document}